\documentclass[twocolumn,superscriptaddress,floatfix,nofootinbib]{revtex4-1}
\usepackage{graphicx}
\usepackage{amsmath}
\usepackage{amssymb}

\graphicspath{{figures/}{.}}

\usepackage[]{hyperref}
\hypersetup{
  colorlinks   = true,
  urlcolor     = blue,
  linkcolor    = blue,
  citecolor   = red
}

\usepackage{mathtools}

\newcommand{\figref}[1]{Fig.~\ref{#1}}

\newcommand{\secref}[1]{Sec.~\ref{#1}}

\renewcommand{\d}[1]{\ensuremath{\operatorname{d}\!{#1}}}

\newcommand{\xpdf}{\psi_{\xi}}

\newcommand{\ftxpdf}{\ft\psi_{\xi}}

\newcommand{\tpdf}{\psi_{\tau}}

\newcommand{\lttpdf}{\lt\psi_{\tau}}

\newcommand{\tsurv}{\Psi_{\tau}}

\newcommand{\lttsurv}{\lt\Psi_{\tau}}

\newcommand{\ilatr}[1]{{\mathcal{L}}^{-1}\left\{{#1}\right\}}

\newcommand{\lt}{\hat}

\newcommand{\ft}{\tilde}

\newcommand{\flt}[1]{{\hat{\tilde #1}}}

\newcommand{\Kr}{{\mathcal K}_{\text{r}}}
\newcommand{\Kd}{{\mathcal K}_{\text{d}}}

\newcommand{\Kra}{{K}_{\text{r}}}
\newcommand{\Kda}{{K}_{\text{d}}}

\newcommand{\Da}{\mathcal D_\alpha}

\newcommand{\ts}{\tau_\text{m}}

\newcommand{\tr}{\tau_\text{r}}

\newcommand{\TD}{S}

\newcommand{\Dt}{\Delta}

\newcommand{\lv}{s}

\newcommand{\rv}{r}

\newcommand{\rdens}{\psi_{\rho}}

\newcommand{\rhohi}{\rdens^*}

\newcommand{\ensav}[1]{\left\langle #1 \right\rangle}

\newcommand{\msd}{m}

\newcommand{\msds}{m_{\text{s}}}

\newcommand{\pp}{P}

\newcommand{\cxt}{{p(x,t| \, t < \TD)}}

\newcommand{\cs}{p_{\text{s}}}

\newcommand{\Ind}[1]{\mathbb I({#1})}

\newcommand{\csicaf}{\affiliation{Spanish National Research Council (IDAEA-CSIC), E-08034 Barcelona, Spain}}
\newcommand{\icfoaf}{\affiliation{ICFO--Institut de Ci\`encies Fot\`oniques, Mediterranean
Technology Park, 08860 Castelldefels, Spain}}

\begin{document}

\newcommand{\figw}{ {1.0\linewidth} }

\title{Reaction-diffusion with stochastic decay rates}

\author{G. John \surname{Lapeyre} Jr.}
\csicaf
\icfoaf
\author{Marco Dentz}
\csicaf

\date{\today}

\begin{abstract}
  Understanding anomalous transport and reaction kinetics due to microscopic
  physical and chemical disorder is a long-standing goal in many fields
  including geophysics, biology, and engineering. We consider reaction-diffusion
  characterized by fluctuations in both transport times and decay rates.  We
  introduce and analyze a model framework that explicitly connects microscopic
  fluctuations with the mescoscopic description. For broad distributions of
  transport and reaction time scales we compute the particle density and derive
  the equations governing its evolution, finding power-law decay of the survival
  probability, and spatially varying decay that leads to subdiffusion and
  an asymptotically stationary surviving-particle density.  These anomalies are clearly
  attributable to non-Markovian effects that couple transport and chemical
  properties in both reaction and diffusion terms.
\end{abstract}

\maketitle

\section{Introduction \label{sec:intro}}

Due to the interaction of diffusion and reaction mechanisms, reaction-diffusion
systems in fluctuating environments may develop collective behaviors that are
very different from those occurring under well mixed conditions. Smoluchowski's
theory~\cite[][]{Smoluchowski1917} quantifies the interaction of diffusion and
reaction for fast bimolecular reactions through an effective rate that is
proportional to the molecular diffusion coefficient. This approach is valid
under well-mixed conditions. Spatial and temporal fluctuations may lead to the
segregation of the reactants~\cite[][]{Ovchinnikov1978} characterized by
non-Poissonian encounter processes and broad first-passage time
distributions~\cite[][]{Benichou2010,Reuveni2010,Reuveni2010a}, such that
reaction kinetics on small and large scales may be very
different~\cite[][]{Avraham2005}.
The sound understanding and quantification of the mechanisms by which
heterogeneity on small scales leads to ``non-classical'' or ``anomalous''
kinetics on large scales plays a central role in applications as diverse as
contaminant degradation and chemical transformations in geological
media~\cite[][]{Steefel2005, Dentz2011a} and chemical kinetics in crowded
intracellular environments~\cite[][]{Schnell2004}.  A number of approaches have
been proposed to model reaction behaviors in heterogeneous environments,
including fractional kinetic orders and time-dependent rate
coefficients~\cite{Kopelman1986, Savageau1995} as well as delayed-reaction
equations \cite{Schnell2004,Bratsun2005,Brett2013,Tian2013}. Oftentimes, such
effective approaches to explain non-exponential survival probabilities are
phenomenology-based lumped parameter models~\cite[][]{Aris1989}.  Indeed, the
variety of mechanisms leading to anomalous diffusion and kinetics precludes
general answers to fundamental questions.  For instance, are emergent anomalous
kinetics better described by non-linear, or by non-Markovian evolution
equations?  We address this question by solving the random decay model for
fluctuations characterized by broad distributions of transport and reaction time
scales, obtaining reaction-subdiffusion equations.  Insisting on an exact
derivation is especially important in this case, since the \textit{prima facie}
reasonable approach of adding reaction terms to known subdiffusion
equations~\cite{Fedotov2002,Mendez2006,Henry2000,Henry2002a,Henry2005,Langlands2007}
has been shown to be inconsistent with microscopic dynamics and
kinetics~\cite{Sokolov2006}.  We find anomalous kinetics associated with
population splitting and identify the cause in non-Markovian, rather than
non-linear effects.  Furthermore, the transport is highly anomalous, with the
particle density approaching a stationary state.

We do not make assumptions regarding the origin of the distribution of
transport times, but rather take these properties as given. However, it is important to
note there do exist derivations in the literature of transport properties, such
as first-passage times, from characteristics of complex media. For instance,
first-passage observables have been computed for diffusion on fractals or media
with heavy-tailed trap distributions~\cite{Benichou2010}, and by applying the
mapping between random walks and vibrations to complex elastic
networks~\cite{Reuveni2010,Reuveni2010a}.

The paper is organized as follows.
In \secref{sec:model} we introduce the random decay model as the continuous time
random walk (CTRW) in which the walker is subject to a random decay process
during each waiting period.
In \secref{sec:GME}, we derive the generalized master equation for the particle
density and derive the solution in Fourier-Laplace space.
In \secref{sec:ratemodel}, we connect the random rate model to a general
stochastic process.  We preview the main results in this connection: the mean
square displacement approaches a constant. The reaction kinetics follow a power
law governed by an evolution equation with a heavy-tailed memory kernel that
couples microscopic transport and reaction parameters.
In \secref{sec:wellmixed}, we show that ordinary diffusion subject to random
decay leads to perfect mixing in the scaling limit, and is equivalent to decay
at a single, average rate.
In \secref{sec:fracRD}, we consider random rates combined with heavy-tailed
waiting times and derive a generalized fractional Fokker-Planck
reaction-diffusion equation with reaction and diffusion kernels that couple
reaction and transport.
In \secref{sec:powerlawrates}, we assume a power-law rate PDF and derive exact
asymptotic expressions for the reaction-diffusion equations and solutions.
In \secref{sec:localization}, we demonstrate localization by deriving the exact
expression for the asymptotic, steady-state particle density as a two-sided
exponential distribution.

\section{The random decay  model\label{sec:model}}

\subsection{Stochastic decay rates  \label{sec:stochasticrates}}

\begin{figure}
    \includegraphics[width=\figw]{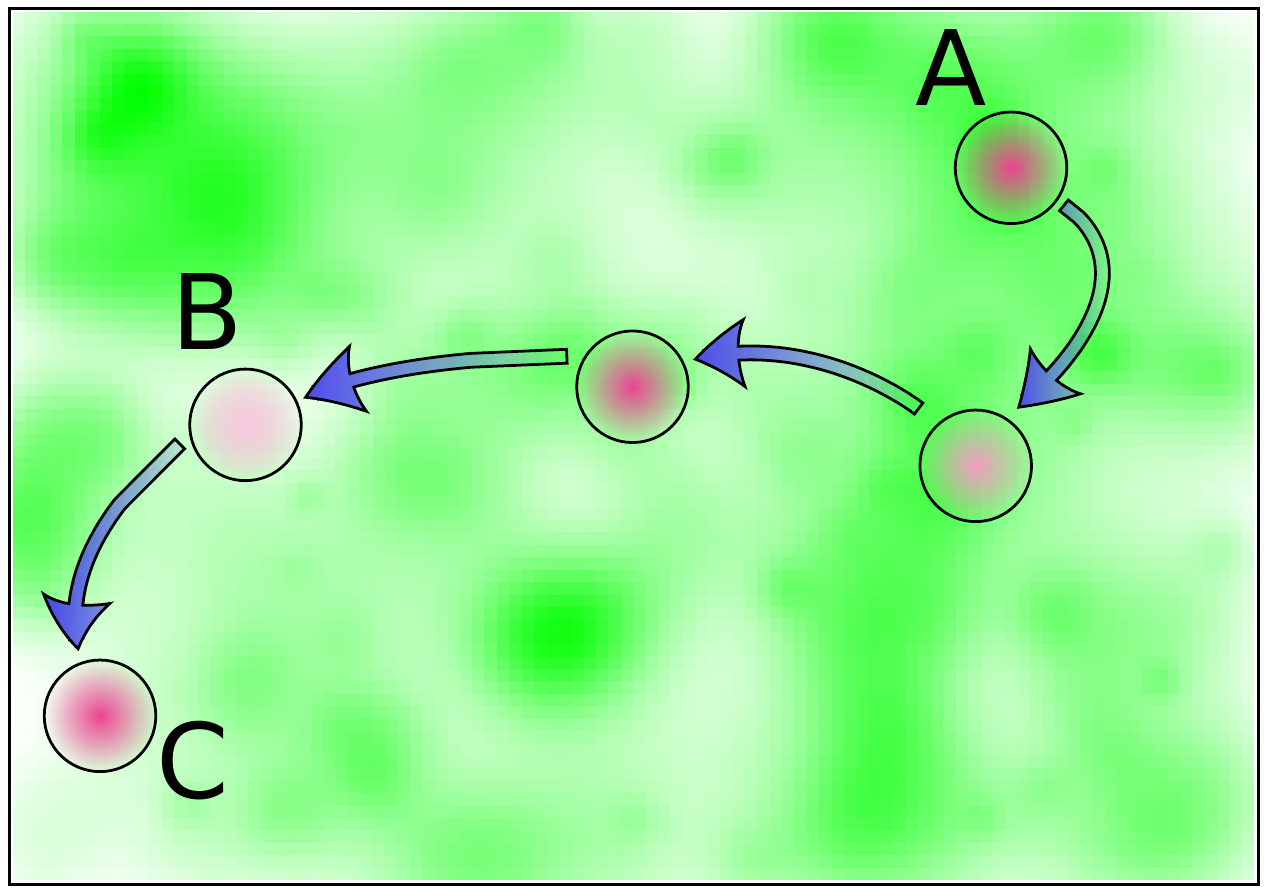}
    \caption{Overview of the model. The particle makes random jumps until it
      decays. Darker green corresponds to faster jumping. Darker red corresponds
      to faster decay. For clarity, the decay rate (red) is shown only at sites
      that the walker occupies. (A) Dark green and dark red: the particle
      experiences a high decay rate for a short time. (B) Light green and light
      red: The particle is immobile for a long time with a small decay rate, and
      so may survive for a long time.  Events like (B) cause anomalous kinetics.
      (C) Light green and dark red: Particle is immobile for a long time with
      high decay rate, and so has a high probability of decaying in this step.
    }
    \label{fig:overview}
\end{figure}
In this section, we formulate our model of diffusion in a fluctuating physical
and chemical environment as a continuous time random walk (CTRW) subject to
random decay. That is, a particle of species $A$ performs a random walk and at
the same time, undergoes an irreversible reaction $A\rightarrow B$.  The
particle waits a random time $\tau$ before making each step, and undergoes decay
at a rate $\rho$ during this waiting period, where $\tau$ and $\rho$ vary
randomly.  Regions where long waiting times and slow decay coincide are
responsible for subdiffusion and anomalous kinetics, as illustrated
in~\figref{fig:overview}.  This type of quenched disorder in the reaction and
diffusion properties may occur in heterogeneous geological media characterized
by a spatial distribution of minerals and thus specific reactive surface, and
porosity~\cite[][]{Steefel2005}, which leads to scale effects in the reaction
properties~\cite[][]{Meile2006,Li2008}.

In this paper, we aim at quantifying the impact of variability in the physical
and chemical system properties on the reaction behavior. To this end, we make
the simplifying assumption that each random waiting time is independent of all
past waiting times, and each decay rate is independent of all past decay
rates. In other words, we assume fully annealed disorder, ignoring possible
effects of correlations between steps.  Models of quenched disorder assume that
the medium fluctuates slowly enough that the walker samples a static
configuration. However, annealed disorder is inherent in many systems in which
the timescales of thermal fluctuations and stochastically varying transport and
kinetic parameters are not well separated.  For example, the waiting times may
arise from long excursions into cul-de-sacs where the duration of the excursion
is due to the motion of the particle itself, and so is memoryless. This feature
has been exploited in studying reaction-subdiffusion front propagation in spiny
dendrites~\cite{Iomin2013}.  Furthermore, the stochasticity may stem from a
fluctuating internal state of a particle that interacts with the medium.
Examples of the latter are conformational fluctuations of
proteins~\cite{Manzo2015} and enzymes~\cite{Boehr2009}.  It is worth noting
that, in distinction from quenched models, annealed models are often solvable.
In some cases, these solutions provide insight into features of the
corresponding quenched model that do not depend strongly on correlations. This
is the case for the model considered here. We derive exact results showing
anomalous kinetics due to memory effects. Preliminary numerical results show
that exponents characterizing power-law behavior of the mean square displacement
~(MSD) and survival probability are the same in annealed and quenched versions of
the model~\cite{Lapeyre2017}, which suggests that the same memory effects are at
play in the quenched setting.

The structure of the annealed model allows the following practical
simplification, and at the same time, generalization.  During the $n$th step,
the particle is subject to two random processes, one triggering spatial
translation and the other the conversion from species $A$ to species $B$. For
example, the waiting time before translation may be due to thermally driven escape from a trap with
random energy. And the reaction rate may depend on the random local
concentration of a catalyst.  Escape from the trap will interrupt the
reaction. And reaction will effectively interrupt escape by removing the
particle from the population whose concentration we are measuring.
%
\newcommand{\rt}{t}
Since one can view the jumping as interrupting an ongoing reaction and starting a new one,
the $n$th step may be simulated as follows.
Sample a waiting time $\tau^{*}$ from the PDF of $\tau$ and
a rate $r$ from the PDF of $\rho$.
Then, sample a random decay time $\rt$ from the PDF
\begin{equation}
 \psi_{r}(\rt|r) = r \exp(-\rt r).
  \label{psir}
\end{equation}
If $\tau^{*} < \rt$, the reaction is interrupted and the particle jumps. On the
other hand, if $\rt\le \tau^{*}$, the particle indeed decays.
But, sampling first a decay rate, and then a decay time is mathematically
equivalent to sampling the decay time directly from the PDF of a single-step
decay time $\Dt$ given by
\begin{equation}
 \psi_{\Dt}(t) = \int_{0}^{\infty} \d{r} \, \psi_{r}(\rt|r)  \psi_{\rho}(r),
  \label{Dtpdfdef}
\end{equation}
where $\psi_{\rho}(r)$ is the PDF of $\rho$.
It is worth noting that this procedure is equivalent to sampling a
decay time from a PDF $\psi_{\Dt}(t)$ that is
\textit{independent} of the step number $n$.
In other words, Poissonian decay with a random parameter (the rate) is
equivalent to non-Poissonian decay described a single PDF. In fact the argument
above also works if 1) $\Dt$ is due to averaging an arbitrary PDF over a random
parameter, and 2) $\tau$ and $\Dt$ are not independent.  In the following, we
derive the basic results in this generality for two reasons. Firstly, the
results may be applied directly to more complicated situations.  For example, we
will suggest below a stochastic Michaelis-Menten scheme defined by coupling
$\Dt$ and $\tau$. Secondly, as we discuss below, results from the study of a
broad class of stochastic processes can be applied directly to the general
formulation of our model~\cite{Pal2017}.

\subsection{General formulation \label{sec:genmodel}}

We consider a molecular entity of type $A$ (the ``particle'') performing a
continuous-time random walk while undergoing random conversion to an entity of
type $B$.  The particle's position $x_n$ and time $t_n$ are given by
\begin{align}
\label{ctrw}
x_{n+1} = x_{n} + \xi_{n+1}, && t_{n+1} = t_{n} + \tau_{n+1},
\end{align}
where $n \geq 0$, $t_{0}=0$, and the initial density, that is the density of
$x_{0}$ is denoted $p_{0}(x)$. Here, $\tau_{n}$ is the random waiting time before the $n$th step,
and $\xi_{n}$ is the step displacement.  We call the time
interval $[t_{n},t_{n}+\tau_{n+1}]$ the $n$th \textit{renewal period}.  The
particle position at time $t$ is then given by
\begin{equation}
 x(t) = x_{n_t},
 \label{ctrwproc}
\end{equation}
where the renewal process $n_t = \max(n|t_n \leq t)$ is the number of steps
(renewal periods) taken by time $t$. With each renewal period, we associate a
single-step random decay time $\Dt_{n}$. The decay time $\Dt_{n}$ may be due to
a disordered Poissonian process as in~\eqref{Dtpdfdef} or some more complicated
chemical or biological process. It represents the time it would take the
particle to decay if it were not subject to transport.  Thus, if the particle is
still alive at time $t_{n}$ and $\tau_{n} < \Dt_{n}$, then the particle takes
the $n$th step before it decays, and thus survives decay. On the other hand, if
$\Dt_{n} \le \tau_{n}$, then the particle decays before it has a chance to take
a step. The random renewal period $m$ during which the particle decays is given
by $m = \min(n| \Dt_{n} < \tau_{n})$. Thus, the random time $\TD$ at which the
particle finally decays after zero or more periods is given by
\begin{equation}
   \TD = \sum_{n=1}^{m-1} \tau_{n} + \Dt_{m}.
  \label{Rrecurs}
\end{equation}
Note that the clock $t_{n}$ tracks only the step waiting times, but not the
decay time.  The analysis is facilitated by this choice, that is, considering an
ordinary CTRW for which we mark the special time $\TD$.  Note also, that this
framework is different from kinetic Monte-Carlo approaches such as the (spatial)
Gillespie method~\cite[][]{Gillespie1977,Kampen2007}, which treats both
diffusive and reactive particle events on the same ground. In the present work, particles
perform a spatial random walk according to~\eqref{ctrw}, and they may react
during the (physical) waiting time with a certain probability as detailed
above. This approach is equivalent to the reaction-diffusion equation for the
species concentration~\cite{Sokolov2006} and to more general non-local reaction
and reaction-diffusion equations as developed in the remainder of the paper.

We assume annealed disorder, so that each renewal period is independent. That
is, $(\xi_{n},\tau_{n},\Dt_{n})$, $n=1,2,\ldots$ are independent and identically
distributed (iid) copies of $(\xi,\tau,\Dt)$.  For the moment, we allow that
$\tau$ and $\Dt$ may be dependent as they may be coupled by chemical and
physical properties of the medium. Although we mention such situations
in~\secref{sec:summary}, for the bulk of the paper, we will assume that they are
uncoupled. For simplicity we assume that the step displacement $\xi$ is
independent of both $\Dt$ and $\tau$ and is distributed according to the
probability density function (PDF) $\xpdf(x)$ satisfying
\begin{align}
\label{xidef}
\ensav{\xi} = 0,  && \ensav{\xi^2} = \delta^{2} < \infty,
\end{align}
where angle brackets denote averages and $\delta$ is a microscopic length scale
characterizing the typical jump length. In this paper we distinguish Laplace
transformed quantities by a hat, and Fourier transformed quantities by a
tilde. The Laplace-conjugate of $t$ is $s$ and the Fourier-conjugate of $x$ is
$k$.

The use of the word \textit{coupled} above refers only to whether random
variables representing microscopic quantities are independent.  Below, we shall
be concerned with whether the mesoscopic transport and reaction are
coupled. There is no direct relation between these two concepts. In fact, we
find mescoscopic coupling in the case that all microscopic random variables are
independent.

It is worth noting that the random decay time $\TD$ takes the form of the
generic first passage time (FPT) under reset~\cite{Reuveni2016,Pal2017}.  Here
the ``passage'' is completion of a reaction (decay) at time $\Dt$, and the reset
time $\tau$ begins the reaction anew.  Thus, the random decay model may be
viewed as FPT under reset coupled with CTRW by identifying the reset time of the
FPT with the waiting time of the CTRW. The identification of the survival time
$\TD$ with FPT under reset allows one to immediately apply general results for
FPT under reset~\cite{Pal2017}, including expressions for the mean FPT and
fluctuations.

The main quantities of interest are the following.  We refer to the probability
that the particle has not decayed up to time $t$ as the \textit{survival}
probability, denoted by
\begin{equation}
  p(t) = \Pr(t<\TD).
 \label{ptdef}
\end{equation}
The evolution of $p(t)$ gives information on the chemical kinetics and reaction
dynamics averaged over the entire system. For example, under a constant decay
rate $\rho_0$, it decays exponentially as $p(t) = \exp(-\rho_0 t)$.  The density
of surviving particles is given by the particle density of the CTRW conditioned
on survival, that is, on $t<\TD$. We denote the density of surviving particles by
\begin{equation}
  \cxt.
 \label{cdef}
\end{equation}
Recall that we use the word ``decay'' as a shorthand for any irreversible
reaction $A \rightarrow B$.  Since we assume that the entities are
non-interacting, one may interpret $\cxt$ as either the probability density for
a single $A$ entity, or as the local concentration or number density of $A$
normalized to one.
In the latter case, $\cxt$ is the profile observed at time $t$ by an imaging
technique that detects species $A$, but not species $B$.
According to Bayes' rule and~\eqref{ptdef}, the joint particle density and
probability of survival is then given by
\begin{equation}
   p(x,t) = \cxt p(t).
 \label{pdef}
\end{equation}
For simplicity, we shall refer to $p(x,t)$ as a ``density''.  Because $\cxt$ is
normalized to one, we have the marginal
\begin{equation}
  {\int\limits_{-\infty}^{\infty}}p(x,t)\d{x}= p(t).
  \label{pxtint}
\end{equation}
Although $p(t)$ and $\cxt$ are the physically relevant quantities, $p(x,t)$ is
more accessible mathematically. Thus, we will calculate $p(x,t)$ and obtain
$\cxt$ via~\eqref{pdef} by dividing by $p(t)$.
The mean square displacement $\msd(t)$, given by
\begin{equation}
  \msd(t) = \int\limits_{-\infty}^{\infty} x^{2} \cxt \d{x},
 \label{mdef}
\end{equation}
measures the spatial extent of the surviving particles.

Finally, it is important to note that we focus on rate PDFs with a finite
probability that the rate is either zero, or arbitrarily close to zero. This is
because the anomalous behaviour is driven by the coincidence of very long
waiting times with very long decay times. In~\secref{sec:cutoff}, we discuss how
the anomalies are modified in the case that the minimum possible rate is greater
than zero.

\section{Generalized master equation and solution \label{sec:GME}}

We derive the generalized master equation~\cite{Klafter2011} and solution for
the random decay model.  The particle density $p(x,t)$ defined in~\eqref{pdef}
satisfies (See \secref{app:inteq}.)
\begin{subequations}
\label{ctrwr}
\begin{align}
p(x,t) = \int_0^t \d{t^\prime} \eta(x,t^\prime) \Phi_{\tau \Dt}(t-t').
 \label{1}
\end{align}

Here $\eta(x,t)$ is the incoming live-particle flux at position $x$ at time $t$.
That is, $\eta(x,t) \d{x}\d{t}$ is the probability that the particle is alive and makes a step between times $t$ and $t + \d{t}$
into the region between $x$ and $x + \d{x}$
And the factor
\begin{equation}
  \Phi_{\tau \Dt}(t-t') \equiv  \Pr(t-t' < \min[\tau,\Dt]),
 \label{Phidef}
\end{equation}
is the probability that the particle has survived both translation (i.e. has not escaped from
a trap)  and decay up to time $t$ during
a renewal period beginning at time $t'$.  Thus, the integral counts all
particles that arrived at $x$ at some time $t'$ in the past and that have
neither jumped away, nor decayed in the time $t-t'$ since arriving.  The flux
$\eta(x,t)$ satisfies the Chapman-Kolmogorov type integral equation
\begin{align}
\eta(x,t) &=  \int\limits_{-\infty}^\infty \d{x'} \int\limits_0^\infty d{t'} \eta(x',t')  \xpdf(x-x') \phi_{\tau\Dt}(t-t')
  + p_{0}(x) \delta(t),
   \label{2}
   \end{align}
\end{subequations}
where
\begin{equation}
  \phi_{\tau\Dt}(t) \equiv \psi_{\tau}(t | t < \Dt) \Pr(t < \Dt),
 \label{psitaudef}
\end{equation}
is the joint probability and probability density to survive both translation and
single-step decay until time $t$ and then to make a translation (jump) at time $t$.
Here $\psi_{\tau}(t | t < \Dt) = \ensav{\delta(t-\tau)| t < \Dt}$ is the
PDF of the waiting time $\tau$ conditioned on waiting time smaller than decay time,  $t<\Dt$.  Eq.~\eqref{psitaudef}
expresses particle balance under reaction losses.  Note that Eqs.~\eqref{1}
and~\eqref{2} have the same structure as the governing equations of the
classical CTRW.  But in fact they do not describe a CTRW and the model cannot be
cast as one. This is due to the presence of reactions, with the result that
$\phi_{\tau\Dt}(t)$ is not a waiting time PDF and $\Phi_{\tau \Dt}(t)$ is not
a translation survival probability.  Instead the decay of the single-step
survival probability $\Phi_{\tau \Dt}(t)$ includes two loss terms representing
translations and decay. Taking the derivative of~\eqref{Phidef} we find
\begin{equation}
  \partial_{t}\Phi_{\tau \Dt}(t) = -\phi_{\tau\Dt}(t) - \phi_{\Dt\tau}(t),
  \label{Psider}
\end{equation}
where
\begin{equation}
  \phi_{\Dt\tau}(t) \equiv \psi_{\Dt}(t|t \le\tau)\Pr( t \le \tau),
  \label{phidectaudef}
\end{equation}
is the joint probability to both survive single-step decay and not make a jump until
time $t$, and then to decay at time~$t$.

The system~\eqref{ctrwr} can be combined into the generalized reaction-diffusion
Master equation~(GME)~(See \secref{app:GME}.)
\begin{align}
&\frac{\partial p(x,t)}{\partial t} = - \int_0^t dt^\prime \Kr(t -t^\prime) p(x,t^\prime) \ +
\nonumber\\
&
  \int_{-\infty}^\infty \d{x^\prime} \int_0^t \d{t^\prime} \Kd(t-t^\prime)\xpdf(x-x^\prime)
  \left[p(x^\prime,t^\prime) - p(x,t^\prime) \right],
\label{GMEr}
\end{align}
where we define the reaction kernel $\Kr(t)$ and the diffusion kernel $\Kd(x,t)$
through their Laplace transforms as
\begin{align}
\label{kernels}
  \lt\Kr(s) = \frac{\lt \phi_{\Dt\tau}(s)}{\lt \Phi_{\tau \Dt}(s)},
  &&
  \lt\Kd(s) = \frac{\lt \phi_{\tau\Dt}(s)}{\lt \Phi_{\tau \Dt}(s)}.
\end{align}
The diffusion kernel quantifies the impact of random decay and waiting times on
the spatial motion, the reaction kernel on the particle survival. As is evident
in the kernels, the reaction and transport processes are intimately coupled.

Integration of~\eqref{GMEr} over space reveals the dynamics that govern the
reaction kinetics,
\begin{align}
\label{kineticEq}
\frac{d p(t)}{d t} = - \int_0^t \d{t^\prime} \Kr(t - t^\prime) p(t^\prime).
\end{align}
Eq.~\eqref{kineticEq} is of central importance.  It expresses the impact of
segregation of the reactants and the different reaction and transport histories
on the overall reaction behavior. The Markov property of the reaction process
that underlies the exponential model breaks down in the presence of distributed
reaction and diffusion rates. It is interesting to note that the Gillespie
method~\cite{Gillespie1977} is a fully Markovian method, in which interreaction
waiting times are exponentially distributed. The non-local nature of
\ref{kineticEq} indicates that this is no longer valid in spatially
heterogeneous systems. This is the subject of ongoing work.

The solution to the GME in Fourier-Laplace space is the generalized
Montroll-Weiss equation for the particle density (See \secref{app:genMW}.)
\begin{equation}
  \flt p(k,s) = \frac{ \lt \Phi_{\tau \Dt}(s) \ft p_{0}(k)}{1- \ftxpdf(k) \lt \phi_{\tau\Dt}(s)}.
 \label{genMW1}
\end{equation}
Note that~\eqref{genMW1} involves $\phi_{\tau\Dt}(t)$ defined
in~\eqref{psitaudef} and $\Phi_{\tau\Dt}(t)$ defined in~\eqref{Phidef}, but not
$\phi_{\Dt\tau}(t)$ defined in~\eqref{phidectaudef}.  This is because both
$\phi_{\tau\Dt}(t)$ and $\Phi_{\tau\Dt}(t)$ characterize the event that decay
does not occur, while $\phi_{\Dt\tau}(t)$ characterizes the event that decay
\textit{does} occur.  This is to be expected, since $p(x,t)$ is the density of
particles that have not decayed.

Setting $k=0$ in Fourier space is equivalent to integrating over $x$ in real
space.  Thus, putting $k=0$ in~\eqref{genMW1} and referring to~\eqref{pxtint},
we obtain the expression for the survival probability
\begin{equation*}
  \lt p(s) = \frac{ \lt \Phi_{\tau \Dt}(s)}{1- \lt \phi_{\tau\Dt}(s)}.
\end{equation*}
The mean survival time $\ensav{\TD}$ is given by $\ensav{\TD} = \lt p(0)$, from which we obtain the simple form
\begin{equation}
  \label{TDavg1}
\ensav{\TD} = \frac{\ensav{\min(\tau,\Dt)}}{\Pr( \Dt < \tau)}.
\end{equation}
As mentioned above, the survival time $\TD$ is formally a FPT under reset. A
simple, alternative derivation of~\eqref{TDavg1} from this viewpoint is found in
Ref.~\cite{Pal2017}.  From~\eqref{TDavg1} we see that $\ensav{\TD}$ increases
with 1) increasing probability of large values of both $\Dt$ and $\tau$, and 2)
increasing probability of $\Dt>\tau$.

\section{Stochastic rates and anomalous kinetics \label{sec:ratemodel}}
\begin{figure}
        \includegraphics[width=\figw]{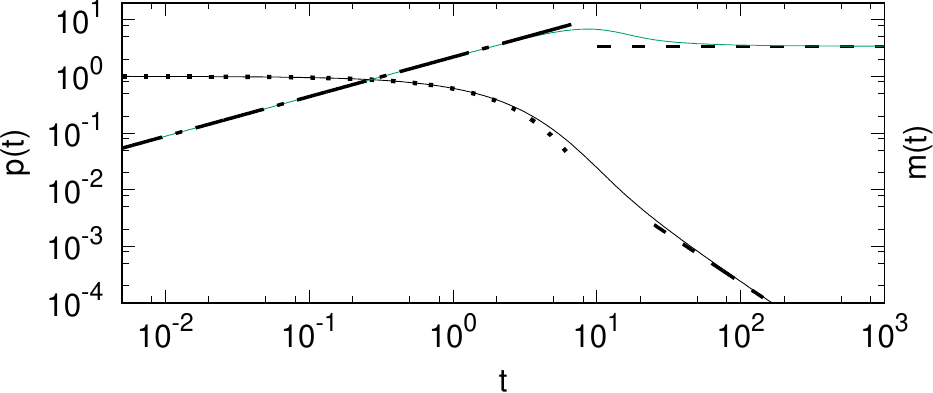}
        \caption{ (Black solid) Survival probability $p(t)$ and (green solid)
          mean square displacement $\msd(t)$ defined in~\eqref{mdef} for
          heavy-tailed waiting time PDF, and power-law reaction rate PDF with
          $\alpha = 0.7$, $\nu = 1$, and $\tr = 1$.  Left and right ordinate
          axes differ in physical dimensions, but are numerically equal.
          (Dotted) Exponential short time behavior of the survival probability,
          which is characterized by the average rate $\ensav{\rho}$.  (Lower
          dashed) Asymptotic power-law decay $p(t) \propto t^{-\gamma}$.
          (Dash-dotted) Short-time power-law behavior
          $\msd(t) \propto t^{\alpha}$.  (Upper dashed) Asymptotically constant
          $\msd(t)$ occurring on the localization time scale
          $\langle \rho^{\alpha} \rangle^{-1/\alpha}$.  }
    \label{fig:PandMSD}
\end{figure}

We now assume that the single-step decay time and the translation waiting time are uncoupled, that is,
$\tau$ and $\Dt$ are independent. We denote the PDF of waiting times $\tau$
by
\begin{equation}
  \tpdf(t) \equiv \ts^{-1}\bar\tpdf(t/\ts),
  \label{psidef}
\end{equation}
where $\bar\tpdf(z)$ is dimensionless and $\ts$ is the waiting time scale.
Furthermore, we adopt the viewpoint of~\secref{sec:stochasticrates} that the
randomness in decay is due to first-order decay with rates that vary
stochastically, but are constant during each renewal period. The variability in
rates is characterized by the PDF
\begin{equation}
  \psi_{\rho}(r) \equiv \tr\bar\rdens(r\tr),
  \label{rdensdef}
\end{equation}
where $\tr$ is the time scale of single-step decay, and $\bar\rdens(z)$ is a dimensionless PDF.

A main result and key message of this paper is that strong physical disorder
expressed via~\eqref{psidef} combined with disordered rates expressed
via~\eqref{rdensdef} leads to anomalous kinetics as well as anomalies in
transport beyond standard subdiffusion.  We quantify these anomalies and
identify their source in long reaction memory rather than nonlinearity.  The
anomalous kinetics and transport are clearly evident in \figref{fig:PandMSD},
which shows the survival probability $p(t)$ and the mean square displacement
$\msd(t)$ for a heavy-tailed waiting time PDF that behaves as
$\tpdf(t) \propto t^{-1-\alpha}$ with $0 < \alpha < 1$, for $t$ larger than the
characteristic time $\ts$ in~\eqref{psidef}, and a rate PDF that behaves as
$\rdens(\rv) \propto r^{\nu - 1}$ with $\nu > 0$ for $\rv$ smaller than the
characteristic rate $1/\tr$.  We observe two remarkable behaviors. Firstly, the
survival probability decays as a power-law $p(t) \propto t^{-\gamma}$, where
\begin{equation}
 \gamma=\alpha+\nu,
 \label{gammadef}
\end{equation}
and secondly, $\msd(t)$ increases proportionally to $t^\alpha$, as for non-reacting particles, until a
characteristic reaction time scale after which it decays towards a
constant. These two behaviors indicate a localization of
the density of surviving particles.

The power-law decay of the survival probability observed in \figref{fig:PandMSD}
can be modeled by a non-linear kinetic rate law as~\cite[][]{Kopelman1986}
\begin{align}
\label{fractal}
\frac{d p(t)}{d t} = - k_e p(t)^{\frac{1 + \gamma}{\gamma}}
\end{align}
with $k_e$ as an effective reaction rate. While this equation gives the
power-law decay $p(t) \propto t^{-\gamma}$, it implies a conceptual framework
that is clearly inconsistent with the correct evolution
equation~\eqref{kineticEq}.  Indeed~\eqref{kineticEq} is linear but
non-Markovian, implying history-dependent evolution.  This is an important point
as the conceptual framework influences the approach taken to more complicated
scenarios. A discussion of its importance in interpreting experiments is found
in Ref.~\cite{Sokolov2012}.

\section{Reaction-diffusion dynamics \label{sec:RDdynamics}}

Spatial fluctuations in biological and physical systems provide our main
motivation for assuming that the random decay has its origin in disordered
rates. Thus, in the following analysis we assume that the single-step decay time
$\Dt$ arises from averaging decay over random rates. However, it may be useful
to go in the opposite direction. Given a distribution for $\Dt$, compute the
corresponding distribution rates. In this way our results, although explicitly
written in terms of random rates, may be applied to random decay times $\Dt$ of
varying physical origin.
The PDF $\psi_{\rho}(r)$ is obtained from that of $\Dt$ as follows.  Referring
to~\eqref{psir}, it is easy to see that the single-step decay-survival
probability is given by
\begin{equation}
   \Pr(t<\Dt) = \ensav{e^{-\rho t}}.
 \label{ratedef}
\end{equation}
Since~\eqref{ratedef} is the Laplace transform of $\rho$, it may be inverted for
any density of $\Dt$ for which the inverse Laplace transform exists.\footnote{%
  An example of a single-step decay-survival probability that does not have a
  Laplace inverse, and thus cannot be expressed via random rates, is a
  deterministic decay time $\psi_{\Dt}(t) = \delta(t-\Dt_{0})$.  }

We begin by writing the GME in terms of random rates.  Using the independence of
$\tau$ and $\Dt$ and referring to~\eqref{ratedef}, we find
$\lt \phi_{\tau\Dt}(s) = \ensav{\lttpdf(s+\rho)}$,
$\lt \phi_{\Dt\tau}(s) = \ensav{\rho\lttsurv(s+\rho)}$, and
$\lt \Phi_{\tau \Dt}(s) = \ensav{\lttsurv(s+\rho)}$, where the
translation survival probability $\tsurv(t)=\Pr(t<\tau)$ is given by
 \begin{equation}
   \tsurv(t)=\int_{t}^{\infty}\tpdf(t')\d{t'}.
  \label{Psidef}
\end{equation}
$\tsurv(t)$ is the probability, in the absence of decay, that the particle has
not taken a step during a renewal period before time $t$.  Thus, the
solution~\eqref{genMW1} to the GME~\eqref{GMEr} may be written as
\begin{equation}
 \begin{aligned}
  \flt p(k,s) &= \frac{\ft p_{0}(k) \ensav{\lttsurv(s+\rho)}}{1- \ftxpdf(k) \ensav{\lttpdf(s+\rho)}}.
 \end{aligned}
 \label{gensolmain}
\end{equation}
Eq.~\eqref{gensolmain} is the basis of the following analysis.  We will describe
the conditions under which on the one hand, the system becomes well-mixed and
exhibits homogeneous kinetics at long times, and on the other hand the system
remains poorly-mixed and exhibits persistent physical and chemical anomalies.
In the following, we assume that the rate density~\eqref{rdensdef} has weight
at, or in the neighborhood of, $r=0$, leaving the more general case
to~Sec.\ref{sec:cutoff}.

The main factors determining the evolution of $p(t)$ and $p(x,t)$, and the
degree of mixing in particular are 1) whether the mean waiting time between
jumps exists, ie $\ensav{\tau} < \infty$.  2) The relative magnitude of the
three time scales: the time scale of microscopic transport $\ts$ defined
in~\eqref{psidef} , the time scale of reactions $\tr$ from~\eqref{rdensdef}, and
the physical time $t$. If $\ensav{\tau} < \infty$, and
\begin{equation}
    \ts<\tr<t,
 \label{timescales}
\end{equation}
then the system tends to a well-mixed state with homogeneous kinetics as the
time scales separate.  This is because at long times surviving particles have
typically made many steps and will sample many rates before dying.  On the other
hand, if $\ensav{\tau}$ diverges, then the system never becomes well mixed, no
matter how large the scale separation in~\eqref{timescales}. Instead, we find
anomalous kinetics and dynamics due to memory effects. This corresponds to
particles that are trapped for long times in regions of low reactivity.

\subsection{Well-mixed scenario.~~ \label{sec:wellmixed}}

We first treat the case $\ensav{\tau}<\infty$.  We consider the scaling limit in
order obtain a mesoscopic picture in which observational length and time scales
are much larger than the microscopic scales.  In the scaling limit
$\delta \to 0$ and $\ts \to 0$ such that
\begin{equation}
\mathcal D = \delta^2 / (2 \ts)
\label{brscalelimit}
\end{equation}
converges to a positive constant, the GME~\eqref{GMEr} reduces to
\begin{equation}
\frac{\partial p(x,t)}{\partial t} = - \ensav{\rho} p(x,t) + \mathcal D \frac{\partial^{2} p(x,t)}{\partial x^{2}},
 \label{ordiffdec}
\end{equation}
provided $\ensav{\rho}<\infty$.
(See~\secref{app:scalinglimit}.)
Eq.~\eqref{ordiffdec} gives a mesoscopic description of evolution of the
system. The local density changes little over a short time, but this time
represents an infinite number of steps.  Of course, physically, displacements
and waiting times may be very small, but must be finite.  Thus,
eq.~\eqref{ordiffdec} is an accurate description insofar as the microscopic and
observational physical scales are well separated. For a colloidal system, the
waiting time is the time between collisions of solvent molecules with a
relatively massive particle, so that the ratio of the time required for the
particle to move a distance equal to its own radius and the time between
collisions may be $6$ orders of magnitude or more.  On the other hand, if the
waiting times are dominated by trapping, then the timescale of the trapping
$\tr$ plays the role of the microscopic time scale and the separation between
$\tr$ and the mesoscopic scale may not be as large. We present an example of the
latter case for $\ensav{\tau}=\infty$ in~\secref{app:MC}.

In the present case,~\eqref{ordiffdec} describes ordinary diffusion with a
constant, homogeneous decay rate.  The survival probability $p(t)$ satisfies the
first-order rate equation
\begin{align}
\frac{d p(t)}{d t} = - \langle \rho \rangle p(t).
 \label{FORE}
\end{align}
The effective reaction kinetics are determined solely by the characteristic
reaction rate. Eq.~\eqref{ordiffdec} makes evident that on the mesoscopic level
the kinetics are effectively homogeneous in space.  Note that this behavior is
also observed in general at times shorter than both the reaction and translation
time scales $\tr$ and $\ts$.  In this case, the reaction kernel also reduces to
$\lt\Kr(s) = \langle \rho \rangle$. This is obtained from~\eqref{kernels} by
considering the limit $s \gg 1/\tr$ and $s \gg 1/\ts$.

The scaling limit leading to~\eqref{brscalelimit} and~\eqref{ordiffdec} involves
letting $\ts$ approach zero. Since we do not rescale the reactions, this implies
$\ts\ll \tr$, which corresponds to a small Damk\"ohler number. Eq.~\eqref{FORE}
immediately gives us the mean survival time $\ensav{\TD} = \ensav{\rho}^{-1}$.
The extreme opposite to the scaling limit is $\tr \ll \ts$ and corresponds to
large Damk\"ohler number.  In this case, the mean survival time is just the mean
single-step decay time $\ensav{\Dt}$. This can be seen by noting that for
$\tr \ll \ts$ it is highly probable that $\Dt < \tau$.  Thus, the numerator
in~\eqref{TDavg1} is approximately $\ensav{\Dt}$ and the denominator
approximately $1$. Furthermore, we note that
$\ensav{\Dt}=\int_{0}^{\infty}\d{t}\Pr(t<\Dt)$, and use~\eqref{ratedef} to
arrive at
\begin{align}
  \ensav{\TD}  \approx \ensav{\Dt} = \ensav{\rho^{-1}}, && \tr \ll \ts,
 \label{timescaleslow}
\end{align}
provided the moment exists.  There is no mixing at all, and $\ensav{\TD}$ is
dominated by particles that never jump, but instead decay in their initial
environments.  The intermediate behavior between these extremes depends strongly
on details of the  distributions of  both the rate and of the waiting-times, rather than just their
asymptotics. We defer a discussion of these more complicated and varied results
to~\secref{supp:noheavytail}.

\begin{figure}
  \includegraphics[width=\figw]{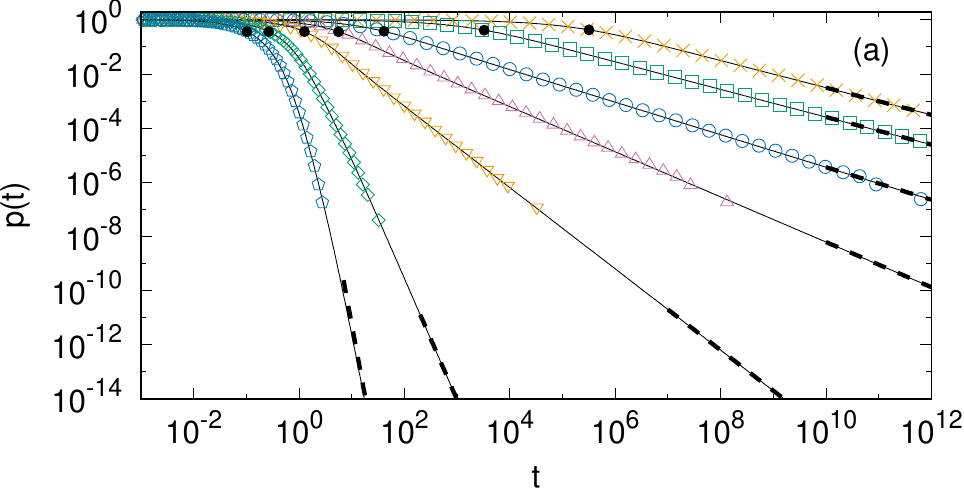}
  \includegraphics[width=\figw]{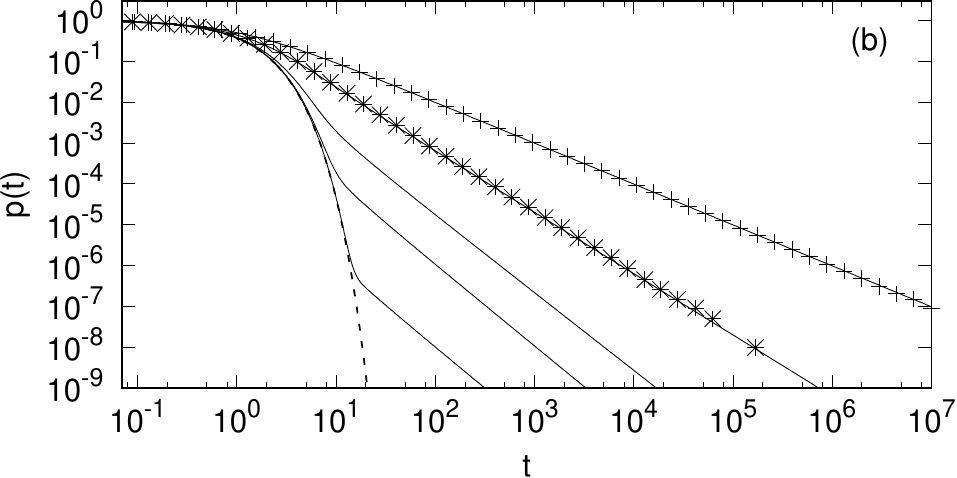}
   \caption{Survival probability $p(t)$ for a heavy-tailed
     $\tpdf(t)$ as in~\eqref{wtdheavy} and a reaction-time PDF
     as in~\eqref{rhopower} with $\tr = 1$.
      (a) $\alpha=1/2$, and (from uppermost to lowermost curve)
    $\nu=10^{-3}, 10^{-2}, 10^{-1}, 1/3, 1, 4, 10$. (Dashed) Asymptotic form~\eqref{totalprobpl}. Filled circles
    indicate the localization time $\tau_l = \ensav{\rho^\alpha}^{-1/\alpha}$.
      (b) $\nu=1$, with (from uppermost to lowermost curve)
     $\alpha=10^{-3}, 1/2, 0.9, 0.99, 0.9999$.
      (dashed) $\exp(- \langle \rho \rangle t)$.
     For both (a) and (b), solid line curves are numerical inversion
    of~\eqref{psurvl},
    symbols are Monte Carlo simulations of the
   microscopic model.
    \label{fig:varyalphanu}
  }
\end{figure}
%

\subsection{Fractional  reaction-diffusion.~~ \label{sec:fracRD}}

As discussed in the Introduction and indicated in Figs.~\ref{fig:overview}
and~\ref{fig:PandMSD}, broad waiting time distributions lead to anomalously long
particle survivals if they coincide with small or vanishing reaction rates. This
inhibits mixing and leads to segregation.  We mentioned above that only in the
case $\ensav{\tau}=\infty$ does this segregation persist in the scaling limit of
vanishing waiting time scale~$\ts$.  We now turn our attention to this
scenario.  We find that the mesoscopic reaction-diffusion equation possesses
kernels that couple the independent microscopic physical and chemical
fluctuations, which manifest the non-Markovian reaction kinetics.
To illustrate, we consider the heavy-tailed waiting time PDF $\tpdf(t)$ that
behaves as
\begin{align}
  \tpdf(t) \sim \frac{\alpha}{\ts\Gamma(1-\alpha)} (t/\ts)^{-1-\alpha}, && 0<\alpha<1,
 \label{wtdheavy}
\end{align}
for times larger than the microscopic time scale $\ts$.  Eq.~\eqref{wtdheavy}
implies that $\ensav{\tau}$ diverges. Physically, this corresponds to waiting
(or trapping) times that occur on all time scales, including the duration of an
experiment. The variation in trapping time may be due to thermal activation over
a random binding energy, or to long, slow, excursions in inclusions, or many
other causes~\cite{Bouchaud1990,Metzler2014}.

The correct scaling limit to employ with~\eqref{wtdheavy} is ${\delta \to 0}$
and ${\ts\to 0}$ such that
\begin{equation}
\Da = \delta^2 / (2 \ts^\alpha)
\label{scalelimit}
\end{equation}
converges to a positive constant.  (See~\secref{app:scalinglimit}.)  In this
limit, the evolution of the particle density is determined by the non-Markovian
reaction-diffusion equation~(See \secref{app:genFFP}.)
\begin{align}
&\frac{\partial p(x,t)}{\partial t} - \frac{\partial}{\partial t}\int_0^t d t' \Kda(t - t^\prime)
\Da \frac{\partial^2 p(x,t^\prime)}{\partial x^2}
\nonumber\\
& = - \frac{\partial}{\partial t} \int_0^t d t^\prime \Kra(t - t^\prime) p(x,t^\prime),
\label{nldr1}
\end{align}
where the reaction and diffusion kernels are defined by their Laplace transforms
\begin{align}
\label{kernelss}
\lt\Kra(s) = \frac{\langle \rho (s + \rho)^{\alpha - 1} \rangle}{s\langle (s + \rho)^{\alpha-1} \rangle}, &&
\lt\Kda(s) = \frac{1}{s\langle (s + \rho)^{\alpha - 1} \rangle}.
\end{align}
For $\rho=0$,~\eqref{kernelss} and~\eqref{nldr1} reduce to the well-known
fractional Fokker-Planck equation.
It is worth noting that the operators in~\eqref{nldr1} describing subdiffusion
with random decay rates may related to fractional calculus via rate-averaged
\textit{tempered fractional calculus}~\cite{Sabzikar2014}.

The solution to~\eqref{nldr1} is  (See~\secref{app:scalinglimit}.)
\begin{equation}
\begin{aligned}
  \flt p(k,s) = \frac{\ensav{(s+\rho)^{\alpha-1}} }
  {\ensav{(s+\rho)^{\alpha}} + k^2 \Da}.
\end{aligned}
\label{mwanomdecay}
\end{equation}
We have assumed here that $p_{0}(x)=\delta(x)$ for simplicity.  The
corresponding survival probability obtained by setting $k=0$ assumes the compact
form
\begin{align}
\lt p(s) = \frac{\langle (s + \rho)^{\alpha - 1} \rangle}{\langle (s + \rho)^{\alpha} \rangle}.
\label{psurvl}
\end{align}
Setting $s=0$ in~\eqref{psurvl}, we see that the mean survival time
$\ensav{\TD}$ of the particle under random diffusion and decay given
by~\eqref{TDavg1} takes the form
\begin{equation}
  \ensav{\TD} = \frac{\ensav{\rho^{\alpha-1}}}{\ensav{\rho^{\alpha}}}.
  \label{mst}
\end{equation}
It is important to note that the scaling limit does not exist if the PDF of the
rates $\rdens(r)$ decays more slowly than $r^{-\alpha-1}$ as $r\to \infty$.  In
this case the denominator of~\eqref{mst} diverges, so that the mean survival
time $\ensav{\TD}=0$.  Likewise, the kernel $\lt\Kra(s)$ in~\eqref{kernelss}
diverges for all $s$, and $\lt p(s)$ in~\eqref{psurvl} is identically
zero. Physically, $\ts\to 0$ means that the rates are sampled very rapidly and
for a heavy-tailed rate PDF there is a high probability of very fast rates.
On the other hand, if $\rdens(r)$ diverges more rapidly than $r^{-\alpha}$ as
$r\to 0$, then the numerator of~\eqref{mst} diverges, so that the mean survival
time $\ensav{\TD}$ diverges. But, in this case, the scaling limit still exists.
For instance, for finite time $t<\infty$,~\eqref{kernelss},~\eqref{mwanomdecay},
and~\eqref{psurvl} are well defined.

\subsection{Broadly distributed mean reaction times.~~ \label{sec:powerlawrates}}
%
In this section, we focus on rate PDFs that decay as a power-law for $r$ much
smaller than the inverse of the characteristic time $\tr$
\begin{align}
  \label{rhopower}
  \rdens(\rv) \sim \frac{\tr \rhohi(\rv \tr)}{\Gamma(\nu)} (\tr \rv)^{\nu-1}, && \nu>0,
\end{align}
where $\lim_{\rv\to 0} \rhohi(\rv) = 1$.  This implies a power-law PDF of the
mean reaction times $\psi_r(t) \propto (t/\tr)^{-1-\nu}$ for $t > \tr$.
Substituting~\eqref{rhopower} into~\eqref{ratedef} we see that the
probability to survive decay in a single step varies asymptotically as (See
\secref{app:pasymp}.)
\begin{equation}
 \Pr(t < \Dt) \sim \left(\frac{t}{\tr}\right)^{-\nu}.
 \label{gdef}
\end{equation}
In general the kernel $\Kra(t)$ approaches the inverse of the mean survival
time~\eqref{mst} at a time comparable to the reaction time scale $\tr$. However,
for power law rates~\eqref{rhopower} and $\gamma<1$, with $\gamma$ given
by~\eqref{gammadef}, computing \eqref{mst} gives $\ensav{\TD}=\infty$,
and~\eqref{kernelss} gives $\Kra(t)\sim t^{\gamma -1}$.  In fact, in this case,
both kernels~\eqref{kernelss} take a particularly simple form,
$\lt\Kra(s) \propto s^{- \gamma}$ and $\lt\Kda(s) \propto s^{-\gamma}$.  Thus,
for $\gamma < 1$,~\eqref{nldr1} becomes the fractional reaction-diffusion
equation~(See \secref{app:fractional}.)
\begin{align}
\frac{\partial p(x,t)}{\partial t} - \mathcal D_\gamma \frac{\partial^{1-\gamma}}{\partial t^{1-\gamma}}
\frac{\partial^2 p(x,t)}{\partial x^2}
 = - k_r \frac{\partial^{1-\gamma}}{\partial t^{1-\gamma}} p(x,t),
\label{nldr}
\end{align}
where $\mathcal D_\gamma \propto \delta^2 / (2 \ts^\alpha \tr^{\nu})$ and
$k_r \propto \langle \rho^\alpha \rangle / \tr^{\nu}$.  Although the microscopic
reactions are first-order, the macroscopic reaction term in~\eqref{nldr} is
non-Markovian with a memory kernel that couples the microscopic transport and
kinetic parameters.  This is made clear in the equation governing the evolution
of the survival probability
\begin{align}
\frac{d p(t)}{d t} = - k_r \frac{\partial^{1-\gamma}}{\partial t^{1-\gamma}} p(t),
 \label{fractkinetics}
\end{align}
which is obtained by integrating~\eqref{nldr} over $x$.

As mentioned earlier, at short times $t < \tr$, the survival probability is
approximately exponential, $p(t) \approx \exp(- \langle \rho \rangle t)$.  In
the case of power-law distributed rates~\eqref{rhopower} we obtain
from~\eqref{psurvl} the explicit, long time solution (See \secref{app:pasymp}.)
\begin{equation}
 \begin{aligned}
 p(t)  & \sim  \frac{t^{-\alpha-\nu}}{\tr^{-\nu}\ensav{\rho^{\alpha}}\Gamma(1-\alpha)}, && \quad \ts \ll \tr \ll t, \quad \nu > 0.
 \end{aligned}
 \label{totalprobpl}
\end{equation}
Eq.~\eqref{totalprobpl} shows that, as anticipated in the definition of the
fractional-order derivative of~\eqref{nldr}, the exponent observed in
\figref{fig:PandMSD} is given by $\gamma = \alpha + \nu$, which manifests again
the intimate coupling of diffusion and reaction mechanisms in the mesoscopic
limit.  \figref{fig:varyalphanu} shows the dependence of $p(t)$ on $\alpha$ and
$\nu$, and the excellent agreement of the derived analytical expressions with
Monte-Carlo simulations of the microscopic model. In Sec~\ref{app:MC} we give a
detailed description of the Monte-Carlo algorithms.

\begin{figure}
         \includegraphics[width=\figw]{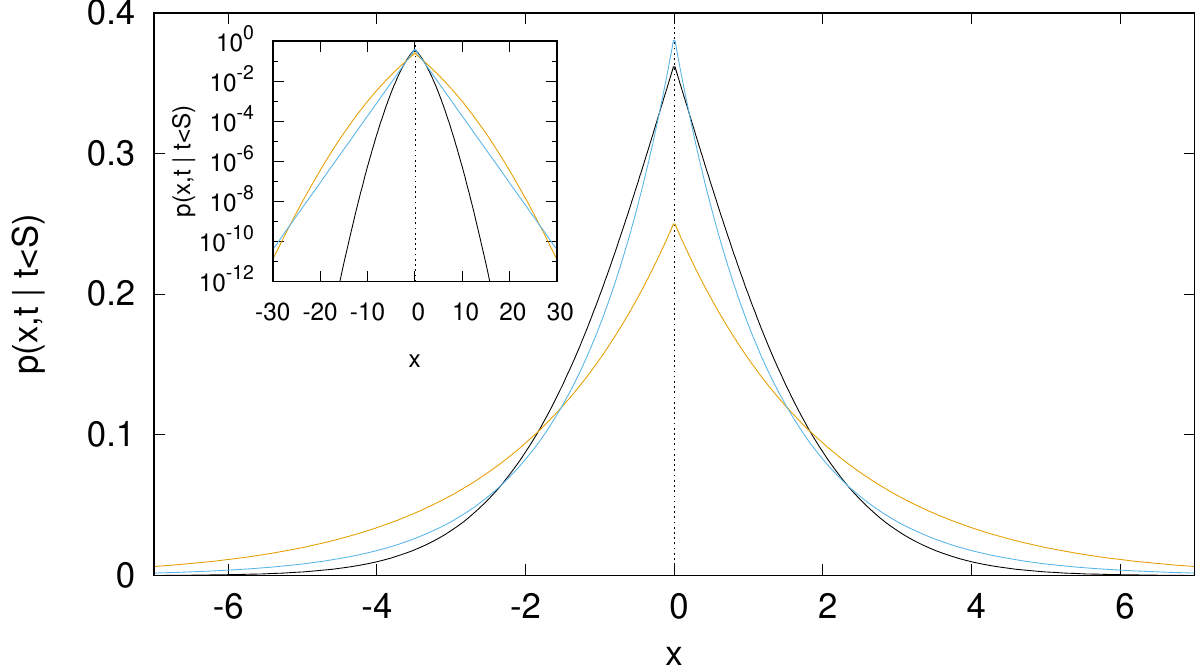}
         \caption{Density of surviving particles $\cxt$ with $\tpdf(\tau)$ and
           $\rdens(\rho)$ as in \figref{fig:PandMSD}, and $t=1$ (black), $t=8.6$
           (yellow), $t\to\infty$ (blue).  The curve for $t\to\infty$ agrees
           with~\eqref{twosided}.  The inset shows the same curves on a semi-log
           scale.  Curves are numerical inversion of~\eqref{propagator}.  }
    \label{fig:propagator}
\end{figure}
%
\begin{figure}
         \includegraphics[width=\figw]{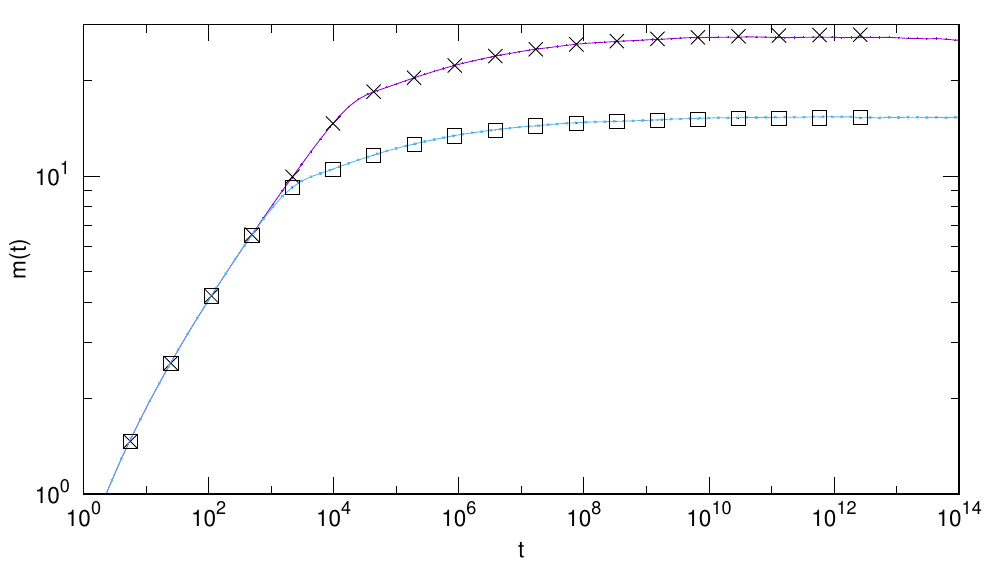}
         \caption{Mean square displacement $\msd(t)$ defined in~\eqref{mdef}.
           Symbols are inverse Laplace transform. Lines are stochastic
           simulations.  Heavy-tailed waiting time PDF~\eqref{wtdheavy} with $\alpha=1/4$,
           $\ts=0.1$. Rate PDF~\eqref{tworates} with $p=1/2$ and (Crosses)
           $\tr=10^{4}$, (Squares) $\tr=10^{3}$.  }
    \label{fig:msdsim}
\end{figure}

\section{Localization of particle density.~~ \label{sec:localization}}

By \textit{localization}, we mean that the surviving-particle density $\cxt$
approaches a stationary density $\cs(x) = \lim_{t\to\infty}\cxt$. This density
has exponential tails and a well-defined, constant, width
$\msds = \lim_{t\to\infty} \msd(t)$.  Thus, a measurement that detects the local
concentration of the surviving species $A$, but not that of the product species
$B$, will be characterized by this width $\msds$. Since the surviving-particle
density $\cxt$ ignores the product species $B$, it is independent of the fate of
$B$, which depends on the application. For instance $B$ may be removed from the
system. Or, it may be invisible to the detector but is either immobilized or
continues to diffuse.  It is interesting to consider the case that species $B$
is immobile, but it \textit{is} detected along with $A$. In this case, the sum
of the local concentrations of $A$ and $B$ approaches the \textit{same}
stationary density $\cs(x)$ obtained by considering species $A$
alone~\cite{Lapeyre2017}.

Localization does not occur in the well-mixed case studied in
Sec.~\ref{sec:wellmixed}.  On the contrary, the decay is spatially uniform. This
is evident by first noting that $p(x,t)=p(t)p_{0}(x,t)$
satisfies~\eqref{ordiffdec} where $p_0(x,t)$ is the particle density for
ordinary diffusion with no decay, ie $\ensav{\rho}=0$.  Then referring
to~\eqref{pdef}, we see that this implies $\cxt = p_0(x,t)$, which means that
the decay is independent of the transport.  Finally, substituting this last
equality into~\eqref{mdef} shows that the MSD $\msd(t)$ evolves exactly as in
the non-reactive case, increasing without bound.  However, in the case of strong
chemical and physical fluctuations, when the system remains poorly-mixed, the
particles are localized at long times.  The surviving-particle density tends to
a stationary state $\cs(x) \equiv \lim_{t\to\infty} \cxt$, given by
(See~\secref{app:scalinglimit}.)
\begin{equation}
 \cs(x) = \frac{1}{2l} e^{-|x|/l},
 \label{twosided}
\end{equation}
where the \textit{localization time} $\tau_{l}$ and \textit{localization length} $l$ are given by
\begin{equation}
   \tau_l = \ensav{\rho^\alpha}^{-1/\alpha} \ \text{ and } \
  l = \sqrt{\Da \tau_l^{\alpha}}.
 \label{localization}
\end{equation}
The corresponding MSD approaches a constant value given by
\begin{equation*}
\msds = \lim_{t\to\infty}\msd(t) = 2 \Da \tau_{l}^{\alpha} = 2l^2.
\end{equation*}
This localization is clearly verified and illustrated in both the MSD $m(t)$ in
~\figref{fig:PandMSD} and~\figref{fig:msdsim}, and the particle density
in~\figref{fig:propagator}.  The MSD approaches a constant at long times. As
$t\to\infty$, the density of surviving particles approaches~\eqref{twosided}
which is represented by the blue curve in~\figref{fig:propagator}.  Note that
the localization time $\tau_l$ marks the scale at which the mean square
displacement crosses over from the power-law behavior $\msd(t) \propto t^\alpha$
to the constant value, as illustrated in \figref{fig:PandMSD}. The deviation of
$m(t)$ from a power-law for $t<\tr$ in~\figref{fig:msdsim} is due to corrections
to the scaling limit.  See~\secref{app:MC} for details of the numerical methods.

To recap, we have derived the fractional reaction-diffusion
equations~\eqref{nldr1},~\eqref{nldr} and fractional kinetic
equation~\eqref{fractkinetics} in the scaling limit of the random walk. These
are exact solutions of the microscopic model with no homogenization or
upscaling.  The presence of memory kernels coupling the transport and kinetic
parameters manifests the poor mixing, even in the scaling limit, in contrast to
the perfect mixing in the scaling limit for Brownian
diffusion~\eqref{ordiffdec}.  We have derived exact expressions in the scaling
limit for the particle density~\eqref{mwanomdecay} and survival
probability~\eqref{psurvl}.  We presented the asymptotic solutions for the
survival probability~\eqref{totalprobpl} and for the localized (stationary)
particle density~\eqref{twosided} and~\eqref{localization}.  These derivations
and their physical interpretation are the main results of~\secref{sec:RDdynamics}
and~\secref{sec:localization}.

\section{Coupled vs. uncoupled reaction.~~ \label{sec:othermodels}}

To better understand stochastic decay, it is useful to compare the mesoscopic
description of the random decay model to that of other models of
reaction-subdiffusion.  We refer to a model in which the reaction proceeds
independently of the transport as ``uncoupled''. Otherwise, it is
``coupled''. The question of whether a model is coupled or uncoupled is an
instructive point of comparison, which we address in the following.

\subsection{Uncoupled reaction}

Suppose $\pp(x,t)$ is the density of surviving particles undergoing subdiffusion
and an unspecified decay process. Define $q(x,t)$ via
\begin{equation}
 q(x,t) \equiv \frac{\pp(x,t)}{p_{0}(x,t)},
  \label{qdef}
\end{equation}
where $p_{0}(x,t)$ is a solution to the fractional Fokker-Planck equation with
no decay~\cite{Metzler2004}
\begin{equation}
\frac{\partial p_{0}(x,t)}{\partial t} = K_{\alpha}  \, D_{t}^{1-\alpha}  \frac{\partial^2 }{\partial x^2} p_{0}(x,t),
  \label{ffpeq}
\end{equation}
and $D_{t}^{1-\alpha}$ is the Riemann-Liouville fractional derivative~\cite{Metzler2004}.
Substituting $p_{0}(x,t) = \pp(x,t)/q(x,t)$ into~\eqref{ffpeq} we see that
$\pp(x,t)$ satisfies the equation
\begin{equation}
  \frac{\partial \pp(x,t)}{\partial t} = q(x,t) K_{\alpha}  \, D_{t}^{1-\alpha}\frac{\partial^2 }{\partial x^2}\left[ \frac{\pp(x,t)}{q(x,t)} \right]
   + \frac{\partial_{t}q(x,t)}{q(x,t)} \pp(x,t).
  \label{ffpuncoup}
\end{equation}
By construction,~\eqref{ffpuncoup} holds formally for any density $\pp(x,t)$,
with $q(x,t)$ given by~\eqref{qdef}.  But it is evidently only meaningful if
$\pp(x,t)$ results from a particle that diffuses according to~\eqref{ffpeq}, and
is subject to decay that is \textit{independent} of the
dynamics~\cite{Sokolov2006,Sagues2008,Abad2010,Fedotov2010,Abad2013a,Yuste2013}.
This becomes clear upon considering the time rate of change of mass at position
$x$ and time $t$
\begin{equation}
 \frac{\partial_{t} \pp(x,t)}{\pp(x,t)}.
  \label{instdecay}
\end{equation}
Using~\eqref{qdef} we write~\eqref{instdecay} as
\begin{equation}
 \frac{\partial_{t} \pp(x,t)}{\pp(x,t)} =  \frac{\partial_{t} p_{0}(x,t)}{p_{0}(x,t)} + \frac{\partial_{t} q(x,t)}{q(x,t)}.
  \label{instdecaytwo}
\end{equation}
The first term on the right hand side is the rate due to transport.  We are
interested in the second term $[\partial_{t} q(x,t)] / q(x,t)$, which is the
instantaneous decay rate (times $-1$) at position $x$.  The role of the second
term as a time and space dependent decay rate is also clear in the last term
in~\eqref{ffpuncoup}.  The diffusion and decay in~\eqref{instdecaytwo} are
manifestly independent.  It is important to note that the reaction term
in~\eqref{ffpuncoup} is Markovian, that is, local in time.  Indeed,
integrating~\eqref{ffpuncoup} over space, we find an equation for the survival
probability
\begin{equation}
    \frac{\partial \pp(t)}{\partial t} = \int\limits_{-\infty}^{\infty} \d{x} \frac{\partial_{t}q(x,t)}{q(x,t)} \pp(x,t).
   \label{survind}
\end{equation}
If we allow $q(x,t)$ to depend on the density $\pp(x,t)$ itself,
then~\eqref{survind} is non-linear~\cite{Fedotov2010}.  Still, the decay is
independent of the dynamics and is Markovian.  Eq.~\eqref{ffpuncoup} has been
derived for many models of uncoupled dynamics and decay, appearing, for example,
as Eq.~(20) in Ref.~\cite{Sokolov2006}, Eq.~(29) in Ref.~\cite{Abad2010}, and
Eq.~(23) in Ref.~\cite{Fedotov2010}.  Typically, $q(x,t)$ is independent of
$\pp(x,t)$, so that any $x$-dependence in $q(x,t)$ represents independent,
externally imposed, spatially varying decay.

\subsection{Coupled reaction}

The random decay model strongly couples chemical kinetics and transport, which
results in a very different description and behavior.  There is no explicit
space-dependent decay in the microscopic model of~\secref{sec:genmodel} as there
is in~\eqref{ffpuncoup}.  However the strong coupling results in a non-Markovian
reaction term in~\eqref{nldr1} and in~\eqref{nldr}, which in turn gives rise to
a time- and space-dependent decay rate $R(x,t)$.  The decay rate $R(x,t)$ is
that part of the time rate of change of the density $p(x,t)$ that is not due to
transport. It is an \textit{effective} or \textit{mesoscale} rate that emerges
from microscopic kinetics and transport that have no explicit space dependence.

To compute $R(x,t)$ for the random rate model, we begin by
dividing~\eqref{nldr1} by $p(x,t)$, thereby obtaining an expression for the time
rate of change of the mass that is analogous to the expression for independent
decay~\eqref{instdecaytwo}.  Then $R(x,t)$ is given by the last term
in~\eqref{nldr1} divided by $p(x,t)$,
\begin{equation}
  R(x,t) = [p(x,t)]^{-1} \frac{\partial}{\partial t} \int\limits_0^t d t^\prime \Kra(t - t^\prime) p(x,t^\prime) .
  \label{instrate}
\end{equation}
$R(x,t)$ depends on the history of the particle density at $x$ through the kernel $\Kra(t)$.
Thus, it is the non-Markovian operator that induces a spatial dependence in the
effective decay rate.
%
\begin{figure}
         \includegraphics[width=\figw]{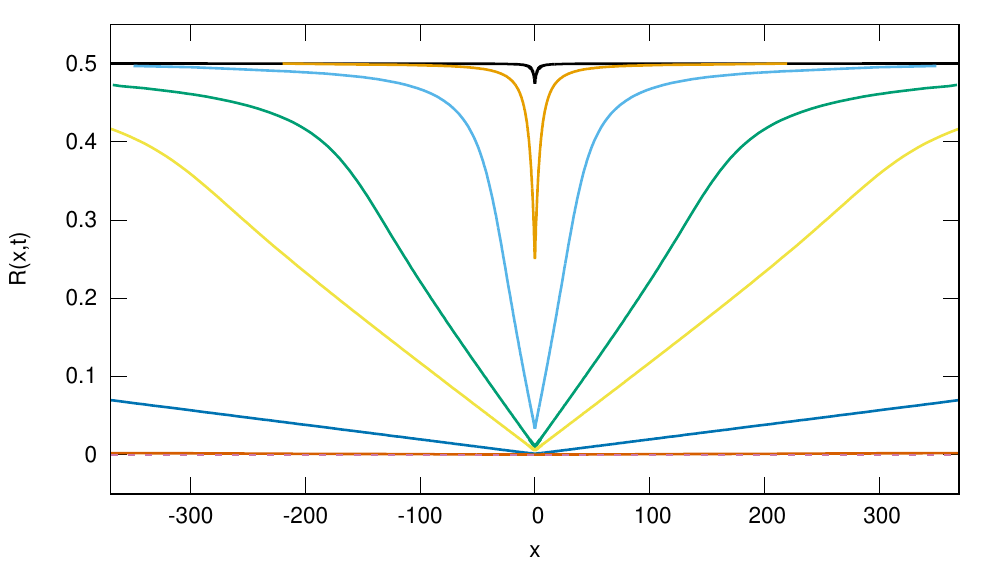}
         \caption{Decay rate $R(x,t)$ defined in~\eqref{instrate}. Times from
           uppermost to lowermost curve (black,gold,light blue,green,yellow,dark
           blue, orange): $t= 0.6, 7, 42, 136, 246, 1450, 5 \times 10^{4}$.
           Waiting time PDF~\eqref{wtdheavy}, $\alpha=0.7$, Decay rate
           PDF~\eqref{uppertrunc}, $\tr = 1$, $\nu = 1$.  Scaling limit with
           generalized diffusivity $\mathcal D_{\alpha} = 1$.  }
    \label{fig:instantrate}
\end{figure}
The solution to~\eqref{instrate} by numerical inversion of the Laplace transform
is shown in~\figref{fig:instantrate}. At short times $t\ll \tr$, the decay is
uniform and exponential with rate $\ensav{\rho} = 1/2$. This corresponds to the
dotted line in~\figref{fig:PandMSD}.  At intermediate
times,~\figref{fig:instantrate} clearly shows a strongly inhomogeneous decay
rate. Because $p(t)$ decays as a power at long times, the instantaneous decay
rate averaged over space $[\partial_{t}p(t)]/p(t)$ decreases like $t^{-1}$. As
$t$ increases, the decay rate near $x=0$ approaches zero, but the asymptotic
value as $x$ approaches $\pm\infty$ is $\ensav{\rho}$. This suppression of the
decay rate in the central part of the density is responsible for the
localization discussed in~\secref{sec:localization}.

Another case of coupling transport and decay is that in which the walker does
not decay while waiting, but rather only before or after making a
step~\cite{Henry2006,Abad2013,Yuste2013}. Suppose a fraction $p$ of walkers are
removed at the beginning of each waiting period. Compare this to the random
decay model with rate density
\begin{equation}
  \psi_{\rho}(r) = (1-p)\delta(r) + p \delta(r-\tr^{-1}),
  \label{tworates}
\end{equation}
which means that during each waiting period the walker suffers no decay with
probability $1-p$ and decays at rate $1/\tr$ with probability $p$.  It can be
shown that~\eqref{nldr} holds in this case with $\nu=0$.  For times $t\gg \tr$,
the longest trapping times are important, so that the particle decays very early
in the waiting period. This is equivalent to removing the walker with
probability $p$ at the beginning of the step. The fractional reaction-diffusion
equation for the latter model given in Ref.~\cite{Henry2006} is indeed equal
to~\eqref{nldr} with $\nu=0$.

\section{Lower cut-off in rate PDF.~~ \label{sec:cutoff}}

Thus far, we have considered rate distributions with rates arbitrarily close to
zero.  But, suppose we shift $\rho$, that is, let $\rho \rightarrow \rho + r_c$
with $r_c > 0$ so that the probability that $\rv<\rv_{c}$ is zero.  We show in
Appendix~\ref{app:shift} that the propagator $p_c(x,t)$ for the shifted reaction
rates is
\begin{align}
\label{pcp}
p_c(x,t) = \exp(-r_c t) p(x,t),
\end{align}
where $p(x,t)$ is the solution for the unshifted density $\rdens(\rv)$.  Note
that this leaves $\cxt$ unchanged, so that $p_c(x,t)$ shows the same
localization as $p(x,t)$.  However, the asymptotic survival
probability~\eqref{totalprobpl} now decays exponentially fast with the
\textit{smallest} rate $r_c$. For instance, for the power law
$\rdens(\rv)$~\eqref{rhopower} and $\rv_{c}\ll \rv_{0}$ the survival probability
follows the truncated power law $p_c(t) \propto \exp(-r_c t) t^{-\alpha-\nu}$.

\section{Conclusions and Outlook \label{sec:summary}}

We derived the mesoscale behavior of a reaction-diffusion system characterized
by microscopically fluctuating transport and reaction kinetics, using the
framework of a continuous time random walk that samples disordered decay rates.
We showed that broadly distributed waiting and reaction times give rise in the
scaling limit to a generalized fractional reaction-diffusion equation with
non-Markovian reaction and diffusion operators both of which are characterized
by intimate coupling of microscopic chemical and physical parameters.  This
equation describes a system that asymptotically remains poorly mixed leading to
power-law kinetics and spatially inhomogeneous reactions.  The resulting decay
is manifest in a particle density whose profile differs radically from that
given by nonreactive subdiffusive CTRW, most notably in a stationary (localized)
particle density at long times.  This is in stark contrast to the case of
ordinary diffusion in the scaling limit, which experiences spatially uniform
decay with a rate equal to the average of the disordered rates.

Understanding of the mechanisms by which mesoscale behavior emerges from
microscopic disorder plays a key role in diverse physical systems. For example,
observed scale effects in reaction laws and decrease in reactivity on large
scales in geological media~\cite{Kump2000,Li2008,Meile2006} can be attributed to
spatial heterogeneity in the chemical and physical medium properties. Sometimes
these behaviors are modeled by empirical non-linear reaction rate
laws~\cite{Peters2006}. Our results show that physical and chemical system
fluctuations are unambiguously attributable to non-Markovian, but linear
kinetics. The segregation of reactions, here a mobile and an immobile species,
in the presence of fluctuating chemical properties, leads to a broad
distribution of effective reaction time scales, which are composed of both
transport and reaction times. The reaction process itself is history dependent,
as expressed by the non-local kinetic rate law~\eqref{kineticEq}.  This new
understanding of the role of chemical and physical fluctuations provides a
systematic way towards quantifying effective large scale reaction behaviors and
scale effects in reactivity in terms of the physical and chemical heterogeneity
of the host medium in natural and engineered media.  Furthermore, the results
derived for first-order decay can be generalized to more complex chemical
reactions under stochastic reaction and transport rates along the lines of the
approach presented in~\cite{Hansen2015}.

We have focused on transport in the presence of random translation times and decay
rates.  However, it is important to point out that the theory presented here is
independent of the specific physical context in which it was developed. We
derived the main results for a general stochastic framework that combines two
processes, CTRW and first passage under restart, by identifying the CTRW
waiting time with the restart time.  Applications and mathematical properties
of CTRW~\cite{Meerschaert2014,Metzler2014} and first passage time (FPT) under
restart~\cite{Pal2017,Reuveni2016} have been studied intensively. But the
fruitful union of these two theories remains nearly unexplored.  Possible
avenues can be found in the many diverse processes that determine statistics of
the step displacement $\xi$~\cite{Appuhamillage2011,Ramirez2013},
waiting time $\tau$~\cite{Benichou2010,Reuveni2010,Reuveni2010a}, and single-step decay
time $\Dt$~\cite{Dentz2011a}.

However, an important class of chemical processes, namely Michaelis-Menten (MM)
reactions~\cite{Kou2005}, require further generalization of FPT under reset.  In
recent years, advances in single-molecule spectroscopy have opened the
possibility of measuring and controlling~\cite{Lomholt2007a} catalysis on the
level of single, or a few, molecules.  This in turn has spurred the development
of stochastic approaches to MM reactions.  These include considering the effects
of fluctuations~\cite{Grima2009,Pulkkinen2015}, internal states of the
enzyme~\cite{Kolomeisky2011}, and non-Poissonian processes.
A stochastic Michaelis-Menten scheme is obtained from the generic FPT under
reset by delaying restart of the process by a random time $T_{\text{on}}$ after
each interruption.  In catalytic reactions, $T_{\text{on}}$ represents the
rebinding time.  In this stochastic formulation, recent theoretical studies have
predicted experimentally accessible~\cite{Roeffaers2006}, counter-intuitive
kinetics by replacing the classical Poissonian processes governing binding,
unbinding, and catalysis times with non-Poissonian
processes~\cite{Wu2002,Reuveni2014,Rotbart2015}.  The importance of extending
this Michaelis-Menten scheme to include heterogeneous catalysis due to a
fluctuating environment has been recognized in recent
experimental~\cite{Janssen2014} and theoretical~\cite{Reuveni2014} work.  An
attractive possibility is to modify the framework presented herein by including
the rebinding time $T_{\text{on}}$.  This immediately yields a Michaelis-Menten
scheme capable of handling heterogeneous catalysis via diffusion following
unbinding events.  The challenge of understanding the interplay of transport and
Michaelis-Menten-like processes in cellular
environments~\cite{Schnell2004,Reuveni2014} is a particularly promising
candidate for such a Michaelis-Menten-CTRW approach, given that macromolecular
crowding in cells may lead to both CTRW-like
subdiffusion~\cite{Metzler2014,Barkai2012} and modified binding
dynamics~\cite{Morelli2011,Guigas2008,Zhou2008}.


\section{Appendix  \label{app:calcs}}

\subsection{Derivation of the integral equations for the propagator  \label{app:inteq}}

Here we derive~\eqref{ctrwr} from the microscopic model given in
Sec.~\ref{sec:model}.  The assumption of an unbiased walk with finite step
variance [See~\eqref{xidef} and~\eqref{phikdef}.], will be employed when passing
to the scaling limit, but does not enter here.  The particle
density~\eqref{pdef} may be written
\begin{align*}
 p(x&,t) =
   \bigg\langle \delta(x - x_{n_t}) \Ind{t<\TD} \bigg\rangle.
\end{align*}
Recall that the indicator function $\Ind{\cdot}$ is $1$ if the argument is true
and $0$ otherwise.  The factor $\Ind{t<\TD}$ may be decomposed as follows.  From
the transition rules~\eqref{ctrw} and~\eqref{Rrecurs} it follows that, at a
given time $t$, a particle that is at position $x_{n}$ has survived until the
turning time $t_{n}$ with probability
$\ensav{\prod_{i=0}^{n_t-1} \Ind{\tau_{i}<\Dt_{i}}}$, and has survived the last
time interval $(t-t_n)$ from the last turning point to the present time with
probability $\ensav{\Ind{t- t_{n_{t}} < \Dt_{n_{t}}}}$.  Referring
to~\eqref{ctrwproc}, this implies that the particle density at time $t$ is given
by
\begin{align*}
 p(x&,t) = \\
  & \bigg\langle \delta(x - x_{n_t}) \prod_{i=0}^{n_t-1} \Ind{\tau_{i}<\Dt_{i}} \Ind{t- t_{n_{t}} < \Dt_{n_{t}}}
   \bigg\rangle,
\end{align*}
where the random variable $n_t = \max(n|t_n \leq t)$ is the number of steps
performed up to time $t$.  We partition the probability space into disjoint
sets, so that the expectation becomes a sum of expectations
\begin{align*}
 p(x&,t) = \\
  &  \sum\limits_{n=0}^\infty \bigg\langle \delta(x - x_{n}) \prod_{i=0}^{n-1} \Ind{\tau_{i}<\Dt_{i}} \Ind{t- t_{n} < \min[\tau_{n}, \Dt_{n}]}
   \bigg\rangle
\end{align*}
The last factor combines the requirements that particle has neither decayed nor
jumped during the increment $t- t_{n}$.  We now separate explicitly the
contributions up to the last turning point at time $t'$ and during the final
resting interval $t-t'$
\begin{align*}
p(x,t) =& \int\limits_{0}^{t} dt' \sum\limits_{n=0}^\infty \Big\langle \delta(x - x_{n})  \delta(t' - t_{n})
     \\ & \times \prod_{i=0}^{n-1} \Ind{\tau_{i}<\Dt_{i}} \Ind{t- t' < \min[\tau_{n}, \Dt_{n}]} \Big\rangle.
\end{align*}
Because the last factor depends only on $\tau_{n}$ and $\Dt_{n}$, which are
independent of the remaining factors, we split the expectation into two factors
obtaining
\begin{align}
  \label{etadefone}
p(x,t) =& \int\limits_{0}^{t} dt' \sum\limits_{n=0}^\infty \Big\langle \delta(x - x_{n})  \delta(t' - t_{n})
     \\ & \times \prod_{i=0}^{n-1} \Ind{\tau_{i}<\Dt_{i}} \Big\rangle \Pr(t- t' < \min[\tau, \Dt]),
\end{align}
where we have also used the fact that the step variables share
a common distribution.  We now define
\begin{align}
\label{etadeftwo}
\eta_n(x,t) =  \left\langle \delta(x - x_{n}) \delta(t-t_n) \prod_{i=0}^{n-1} \Ind{\tau_{i}<\Dt_{i}} \right\rangle,
\end{align}
and
\begin{align*}
\eta(x,t) = \sum\limits_{n=0}^\infty \eta_n(x,t).
\end{align*}
Eq.~\eqref{etadeftwo} denotes the joint probability density for a particle to
arrive at position $x$ at time $t$ on the $n$th step. With these definitions,
\eqref{etadefone} is rewritten
\begin{align}
 \label{ctrwrone}
p(x,t) = \int\limits_0^{t} d t' \eta(x,t')   \Pr(t- t' < \min[\tau, \Dt])
\end{align}
We analyze $\eta_{n}(x,t)$ by writing $\eta_{n+1}(x,t)$ in the following form
\begin{multline*}
 \eta_{n+1}(x,t) = \int\limits_{-\infty}^{\infty} d x' \int\limits_0^t d t'
    \Big\langle \delta(x' - x_{n}) \delta(t'-t_n)  \\
   \times  \prod_{i=0}^{n-1} \Ind{\tau_{i}<\Dt_{i}}  \delta(x-x'-\xi_n)
  \delta(t-t'-\tau_n) \\
 \times  \Ind{\tau_{n}<\Dt_{n}} \Big\rangle.
\end{multline*}
That this is indeed the expression for $\eta_{n+1}(x,t)$ can be seen by
performing the integrals and eliminating either one of the delta functions for
$x'$ and either one for $t'$, and using~\eqref{ctrw}.
Note that the only random variables appearing in the last three factors in the
expectation are $\xi_n$, $\tau_n$ and $\Dt_n$, while the first three factors
depend only on random variables for $i<n$.  The last three factors are thus
independent of the first three and we can again factor the expectation.
Furthermore, per the Dirac delta $\delta(t - t' - \tau_n)$,
we have $\tau_n = t - t'$.  Thus, we can write
\begin{multline*}
\eta_{n+1}(x,t) = \int\limits_{-\infty}^{\infty} d x' \int\limits_0^t d t' \\
   \Bigg\langle \delta(x' - x_{n})  \delta(t'-t_n)  \prod_{i=0}^{n-1} \Ind{\tau_{i}<\Dt_{i}}
    \Bigg\rangle \\
  \times  \left\langle \delta(x-x'-\xi_n)\right\rangle \left\langle
  \delta(t-t'-\tau_n) \Ind{\tau_{n}<\Dt_{n}} \right\rangle.
\end{multline*}
Now, referring to ~\eqref{etadeftwo}, we identify the first factor in angular brackets with
$\eta_n(x',t')$, the second factor with $\xpdf(x-x')$ and the third factor
with $\phi_{\tau\Dt}(t-t')$. Thus, we obtain
\begin{align}
\label{CK}
\eta_{n+1}(x,t) &= \int\limits_{-\infty}^\infty dx' \int\limits_0^t dt' \xpdf(x-x')  \\ \nonumber
  & \times
   \phi_{\tau\Dt}(t-t') \eta_{n}(x',t').   \\ \nonumber
\end{align}
Summation over $n$ from $0$ to infinity on both sides of~\eqref{CK} gives for
$\eta(x,t)$
\begin{align}
 \label{ctrwrtwo}
 \eta(x,t) &= p_{0}(x) \delta(t) +  \int\limits_{-\infty}^{\infty} d{x'}  \int\limits_0^\infty d{t'} \eta(x',t') \\ \nonumber
   & \times \xpdf(x-x') \phi_{\tau\Dt}(t-t'),  \\ \nonumber
\end{align}
where we have used
\begin{align}
\sum\limits_{n=0}^\infty \eta_{n+1}(x,t)  = \sum\limits_{n=0}^\infty \eta_{n}(x,t) - \eta_{0}(x,t).
\end{align}
Integral equations~\eqref{ctrwrone} and~\eqref{ctrwrtwo}, appear in the main
body of the paper as~\eqref{ctrwr}.

\subsection{Derivation of the generalized Master equation \label{app:GME}}

We derive the generalized Master equation~\eqref{GMEr}. To this end, we Laplace
transform~\eqref{ctrwrone} and~\eqref{ctrwrtwo}, which gives
\begin{align}
    \label{eqmd:1}
\lt p(x,s) &=  \lt\eta(x,s) \lt\Phi_{\tau \Dt}(s)
\\
\label{eqmd:2}
\lt\eta(x,s) & = p_{0}(x) + \int\limits_{-\infty}^{\infty} d{x'}
                 \lt\eta(x',s)
   \ftxpdf(x-x') \lt\phi_{\tau\Dt}(s).
\end{align}
We now solve~\eqref{eqmd:1} for $\lt\eta(x,s)$ and insert it into~\eqref{eqmd:2}
to obtain
\begin{multline}
  \label{eqmd:3}
\lt p(x,s)\frac{1}{ \lt\Phi_{\tau \Dt}(s)} =   p_{0}(x) \\
 + \int\limits_{-\infty}^{\infty} d{x'}  \lt p(x',s)
\frac{\lt\phi_{\tau\Dt}(s)}{\lt\Phi_{\tau \Dt}(s)}  \xpdf(x-x').
\end{multline}
We now rewrite the left hand side tautologically as
\begin{align}
 \nonumber
  s & \lt p(x,s) +  \lt p(x,s) \left[\frac{1}{\lt\Phi_{\tau \Dt}(s)} -s \right]
\\ \nonumber
  = {}  &
  s  \lt p(x,s) +  \lt p(x,s) \left[\frac{1 -s \lt\Phi_{\tau \Dt}(s) }{\lt\Phi_{\tau \Dt}(s)} \right]
\\ \nonumber
\end{align}
Using $-\partial_{t}\Phi_{\tau \Dt}(t) = \phi_{\tau\Dt}(t) + \phi_{\Dt\tau}(t)$
[See~\eqref{Psider}.] we rewrite the numerator, obtaining
\begin{align*}
  s  \lt p(x,s) +  \lt p(x,s) \left[\frac{\lt\phi_{\tau\Dt}(s) +  \lt\phi_{\Dt\tau}(s) }{\lt\Phi_{\tau \Dt}(s)} \right]
\end{align*}
Replacing the left hand side of~\eqref{eqmd:3} with the last expression and rearranging, we obtain
\begin{multline}
 \label{eqmd:8}
s \lt p(x,s) - p_{0}(x) =   - \lt p(x,s) \frac{\lt\phi_{\Dt\tau}(s) }{\lt\Phi_{\tau \Dt}(s)}  \\
  + \int\limits_{-\infty}^{\infty} d{x'}  \left[\lt p(x',s) - p(x,s) \right]
\frac{\lt\phi_{\tau\Dt}(s)}{\lt\Phi_{\tau \Dt}(s)}  \xpdf(x-x').
\end{multline}
Inverse Laplace transform of the last equation gives the generalized Master
equation~\eqref{GMEr} with kernels defined via~\eqref{kernels} in the main body
of the paper.

\subsection{Fourier-Laplace solutions \label{app:montrollweiss}}

\subsubsection{Generalized Montroll-Weiss equation \label{app:genMW}}

Taking the Fourier transform of~\eqref{eqmd:1} and~\eqref{eqmd:2} we obtain
\begin{align}
    \label{eqjl:1}
\flt p(k,s) &=  \flt\eta(k,s) \lt\Phi_{\tau \Dt}(s)
\\
\label{eqjl:2}
\flt\eta(k,s) & = \ft p_{0}(k) +  \flt\eta(k,s) \ftxpdf(k) \lt\phi_{\tau\Dt}(s).
\end{align}
Solving~\eqref{eqjl:2} for $\flt\eta(k,s)$ and substituting the solution
into~\eqref{eqjl:1}, we obtain the generalized Montroll-Weiss
equation~\eqref{genMW1}.

\subsubsection{Shift of the random reaction rate by a constant $r_c$ \label{app:shift}}
%
We consider the effect of shifting the random reaction rate
$\rho \rightarrow r_c + \rho$. Inserting this shift into~\eqref{gensolmain}
gives
\begin{equation*}
 \begin{aligned}
  \flt p_{c}(k,s) &= \frac{\ft p_{0}(k) \ensav{\lttsurv(s+\rho+r_{c})}}{1- \ftxpdf(k) \ensav{\lttpdf(s+\rho+r_{c})}}.
 \end{aligned}
\end{equation*}
Inverse Fourier-Laplace transform gives, by using the shift theorem of the
Laplace transform
\begin{align*}
  p_c(x,t) &= \exp(-r_c t) p(x,t)
\end{align*}
Integrating over $x$ gives
\begin{align*}
  p_c(t) &= \exp(-r_c t) p(t),
\end{align*}
where $p(x,t)$ and $p(t)$ are propagator and survival probability for $r_c =
0$. This derives~\eqref{pcp}.

\subsubsection{Mean square displacement}

The mean square displacement of the surviving particles is given by
\begin{align}
  \label{m}
  \msd(t)  = & - \left.\frac{\partial^2 \ln \ft p(k,t)}{\partial k^2}\right|_{k = 0}
\\[5 pt] \nonumber
     = & -  \ft p(0,t)^{-1} \left.\frac{\partial^2 \ft p(k,t)}{\partial k^2}\right|_{k = 0}.
\end{align}
Under the assumption $p_{0}(x)=\delta(x)$ and that the moments of the random
displacement $\xi$ satisfy~\eqref{xidef}, we obtain from~\eqref{gensolmain} the
explicit Laplace-space expression
\begin{equation}
 \label{exactsecondmom}
 \left.  -\frac{\partial^{2} \flt p(k,s)}{\partial k^2}\right|_{k=0} =
         \frac{\delta^{2}\left\langle \frac{1 - \lttpdf(s + \rho)}{s + \rho} \right\rangle \ensav{\lttpdf(s + \rho)} }
       {(1-\langle \lttpdf(s + \rho) \rangle)^{2}}.
\end{equation}
We obtain the Laplace transform of the moments $\ensav{x(t)^n}$ in the scaling
regime directly from~\eqref{mwanomdecay} as
\begin{multline}
\label{unnormedmom}
(-i)^n \left.\frac{\partial^n \flt p(k,s)}{\partial k^n}\right|_{k = 0}
=  n!{\Da}^\frac{n}{2} \lt p(\lv) \left[\ensav{(s + \rho)^{\alpha}}\right]^{-\frac{n}{2}},\\
  \quad n = 0,2,4,\ldots.
\end{multline}
%

\subsection{Scaling limit \label{app:scalinglimit}}

We first consider the case $\ensav{\tau}<\infty$. This implies the Laplace
transform of $\tpdf(t)$ is $\lttpdf(s) = 1 - s \ts + o(\ts)g(s)$. And
from~\eqref{xidef} the Fourier transform of the step PDF is
\begin{equation}
 \ftxpdf(k) =  1 - \frac{(\delta k)^{2}}{2} +  o(\delta^{2})h(k).
 \label{phikdef}
\end{equation}
The ``o'' notation means $\lim_{\ts\to0} o(\ts)/\ts =0$ and
$\lim_{\delta\to0} o(\delta^{2})/\delta^{2} =0$.  In Fourier-Laplace space, the
particle density $p(x,t | t< \TD)$ is given by~\eqref{gensolmain}
\begin{equation}
 \begin{aligned}
  \flt p(k,s) &= \frac{\ensav{\lttsurv(s+\rho)}}{1- \ftxpdf(k) \ensav{\lttpdf(s+\rho)}},
 \end{aligned}
 \label{app:gensolmain}
\end{equation}
where we take $p_{0}(x)=\delta(x)$ for convenience.  Using the expansions of the
PDFs and noting that~\eqref{Psidef} implies $\lttsurv(s)=[1-\lttpdf(s)]/s$, we
have
\begin{equation}
  \flt p(k,s) = \frac{\ts +  o(\ts)\ensav{\frac{g(s+\rho)}{s+\rho}}}
  {\frac{k^{2}\delta^{2}}{2} + \ts\ensav{s+\rho} + o(\delta^{2})h(k) + o(\ts)\ensav{g(s+\rho)}}.
  \label{app:gjl:1}
\end{equation}
Dividing numerator and denominator by $\ts$ and taking the scaling limit $\delta \to 0$ and $\ts \to 0$ such that
\begin{equation*}
\mathcal D = \delta^2 / (2 \ts)
\end{equation*}
is a positive constant,~\eqref{app:gjl:1} becomes
\begin{equation}
  \flt p(k,s) = \frac{1}
  {\frac{k^{2}\delta^{2}}{2\ts} + s + \ensav{\rho}}.
  \label{app:gjl:2}
\end{equation}
Eq.~\eqref{app:gjl:2} is the Fourier-Laplace transform of the propagator for
Brownian motion with no reactions, with $s$ replaced by $s+\ensav{\rho}$. Thus,
the solution is $p(x,t)=\exp(-\ensav{\rho}t)p_{0}(x,t)$.

For the waiting time PDF~\eqref{wtdheavy}, we have $\ensav{\tau}<\infty$.  Thus,
the Laplace transform of $\tpdf(t)$ is
$\lttpdf(s) = 1 - (s \ts)^\alpha + o(\ts^{\alpha})g(s)$.  Together
with~\eqref{phikdef}, this yields the joint PDF
$\flt\psi(k,s) = 1 - \delta^{2} k^2/2 - (s \ts)^\alpha + o(\ts^{\alpha})g(s) +
o(\delta^{2})h(k)$.  Substituting these asymptotic forms
into~\eqref{app:gensolmain}, we find
\begin{equation}
  \flt p(k,s) = \frac{\ts^{\alpha}\ensav{(s+\rho)^{\alpha-1}} +  o(\ts^{\alpha})\ensav{\frac{g(s+\rho)}{s+\rho}}}
  {\frac{k^{2}\delta^{2}}{2} + \ts^{\alpha}\ensav{(s+\rho)^{\alpha}} + o(\delta^{2})h(k) + o(\ts^{\alpha})\ensav{g(s+\rho)}}.
  \label{app:gjl:3}
\end{equation}
Dividing by $\ts^{\alpha}$ and taking the limit $\delta \to 0$ and $\ts \to 0$
such that
\begin{equation*}
\Da = \delta^2 / (2 \ts^\alpha)
\end{equation*}
converges to a positive constant, we obtain
\begin{equation}
\begin{aligned}
  \flt p(k,s) = \frac{\ensav{(s+\rho)^{\alpha-1}} }
  {\ensav{(s+\rho)^{\alpha}} + k^2 \Da},
 \label{app:mwanomdecay}
\end{aligned}
\end{equation}
which is~\eqref{mwanomdecay}.

The inverse Fourier transform of~\eqref{app:mwanomdecay}
yields the propagator in Laplace space
\begin{align}
\label{propagator}
\lt p(x,s) = \frac{\lt p(s)}{2} {\mathcal R}(s) e^{- |x| {\mathcal R}(s)},
\end{align}
where ${\mathcal R}(s) \equiv \Da^{-1/2} \ensav{(s+\rho)^{\alpha}}^{1/2}$.  As
$s\to 0$ the inverse Laplace transform of~\eqref{propagator} is dominated by the
factor $\lt p(s)$ so that $\lt p(x,s)$ can be approximated by
\begin{equation}
\lt p(x,s) \approx
\lt p(s) \frac{{\mathcal R}(0)}{2} e^{- |x| {\mathcal R}(0)}
 \label{asymppropagator}
\end{equation}
Taking the inverse Laplace transform and dividing by $p(t)$ we obtain the
stationary density $\cs(x)$ in~\eqref{twosided}. On the other hand, for
$t < \tr$, the survival probability is approximately exponential and thus the
propagator $p(x,t) \approx \exp(-\langle \rho \rangle t) p_0(x,t)$, where
$p_0(x,t)$ is the density for non-reactive particles obtained by setting
$\rho = 0$ in~\eqref{propagator}.

\subsubsection{Derivation of generalized fractional reaction diffusion equation \label{app:genFFP}}

We begin with the generalized Montroll-Weiss equation~\eqref{mwanomdecay}, but
include the general initial particle density $p_{0}(x)$.
\begin{equation*}
\begin{aligned}
  \flt p(k,s) = \frac{\ft p_{0}(k)\ensav{(s+\rho)^{\alpha-1}} }
  {\ensav{(s+\rho)^{\alpha}} + k^2 \Da}.
\end{aligned}
\end{equation*}
Multiplying by the denominator, we have
\begin{equation}
\begin{aligned}
  & \flt p(k,s) \ensav{(s+\rho)^{\alpha}}  \\
  & = -\Da k^{2} \flt p(k,s) + \ft p_{0}(k) \ensav{(s+\rho)^{\alpha-1}}
\end{aligned}
 \label{eqkern1}
\end{equation}
We rewrite the  factor on the left hand side as
\begin{equation*}
 \ensav{(s+\rho)^{\alpha}} = s\ensav{(s+\rho)^{\alpha-1}} + \ensav{\rho(s+\rho)^{\alpha-1}},
\end{equation*}
and divide the equation by $s\ensav{(s+\rho)^{\alpha-1}}$ to obtain
\begin{equation}
\begin{aligned}
  & \flt p(k,s) + \flt p(k,s) \frac{\ensav{\rho (s+\rho)^{\alpha-1}}}{s\ensav{(s+\rho)^{\alpha-1}}} \\
  & =  -\Da k^{2} \frac{1}{s\ensav{(s+\rho)^{\alpha-1}}} + \frac{\ft p_{0}(k)}{s}.
\end{aligned}
\end{equation}
This can be written
\begin{equation}
\begin{aligned}
  & \flt p(k,s) +  \lt \Kra(s) \flt p(k,s) \\
  & =  -\Da k^{2} \lt \Kda(s) \flt p(k,s) + \frac{\ft p_{0}(k)}{s}.
\end{aligned}
\end{equation}
where $\Kra(s)$ and $\Kda(s)$ are given by~\eqref{kernelss}.
Inverting we obtain
\begin{align}
&p(x,t) - \int_0^t d t' \Kda(t - t^\prime)
\Da \frac{\partial^2 p(x,t^\prime)}{\partial x^2}
\nonumber\\
& = - \int_0^t d t^\prime \Kra(t - t^\prime) p(x,t^\prime) + p_{0}(x).
\label{app:nldrinteg}
\end{align}
Taking the derivative of~\eqref{app:nldrinteg} gives~\eqref{nldr1}.

\subsubsection{Asymptotic form of $p(t)$ \label{app:pasymp}}

The expression for the survival probability in the scaling limit is~\eqref{psurvl}
\begin{align}
\lt p(s) = \frac{\langle (s + \rho)^{\alpha - 1} \rangle}{\langle (s + \rho)^{\alpha} \rangle}.
\label{app:psurvl}
\end{align}
We begin with the numerator. Exchanging the order of the integrals, we have
\begin{equation*}
  \ilatr{\ensav{(s+\rho)^{\alpha-1}}} = \ensav{\ilatr{s^{\alpha-1}} e^{-\rho t}},
\end{equation*}
from which we obtain
\begin{equation}
  \ilatr{\ensav{(s+\rho)^{\alpha-1}}} = \frac{t^{-\alpha}}{\Gamma(1-\alpha)} \ensav{e^{-\rho t}}.
  \label{jl:3}
\end{equation}
It is easy to show that the numerator in~\eqref{app:psurvl} is more singular
than the denominator for $s\to 0$. This can be seen, for instance, by taking
derivatives until the leading order term diverges as $s\to 0$. So, we can set
$s=0$ in the denominator to find
\begin{equation*}
   p(t)   \sim \frac{t^{-\alpha} \ensav{e^{-\rho t}}}{\ensav{\rho^{\alpha}} \Gamma(1-\alpha)}.
\end{equation*}
Using the density~\eqref{rhopower}, we have
\begin{equation}
  \ensav{e^{-\rho t}} = \frac{\tr^{\nu}}{\Gamma(\nu)} \int_{0}^{\infty} e^{-tr} r^{\nu-1} \rhohi(\tr r) \d{r}
 \label{jl:4}
\end{equation}
Substituting $z=tr$, we find for  $t\gg \tr$
\begin{equation}
  \ensav{e^{-\rho t}} \sim \left(\frac{t}{\tr}\right)^{-\nu}.
 \label{rateav}
\end{equation}
 Substituting~\eqref{rateav} into~\eqref{jl:4}, we find
\begin{align*}
 p(t)  & \sim  \frac{t^{-\alpha-\nu}}{\tr^{-\nu}\ensav{\rho^{\alpha}}\Gamma(1-\alpha)}, &&  \quad \nu > 0,
\end{align*}
which is Eq.~\eqref{totalprobpl} in the main text.

\subsubsection{Fractional reaction-diffusion equation \label{app:fractional}}

We now show that using the power-law rate PDF~\eqref{rhopower} and assuming
$0 < \alpha + \nu < 1$, the operators in~\eqref{nldr1} reduce to standard
Liouville fractional derivatives, which lead to a fractional reaction-diffusion
equation. Although the equation involves only the standard Liouville fractional
derivatives, the order of the derivative depends on \textit{both} reaction and
transport exponents.  The fractional nature of the kernels comes from their
asymptotic divergence as $s\to 0$.  We begin by analyzing the denominator of the
kernels~\eqref{kernelss} $s\langle (s + \rho)^{\alpha - 1} \rangle$.  Using the
rate density~\eqref{rhopower}, the second factor becomes
\begin{align*}
  \langle{(s+\rho)^{\alpha-1}}\rangle & \sim \frac{\tr^{\nu}}{\Gamma(\nu)} \int_0^\infty (s+\rv)^{\alpha-1} \rv^{\nu-1} \rhohi(\rv\tr)  \d{\rv}
\\
  = & \frac{\lv^{\alpha+\nu-1}}{\tr^{-\nu} \Gamma(\nu)}  \int_0^\infty (1+z)^{\alpha-1} z^{\nu-1} \rhohi(\lv z\tr) \d{z},
\end{align*}
where we have made the substitution $r=sz$.

For $s \tr \ll 1$, the last integral converges while the factor
$\rhohi(\lv z\tr)$ remains near $1$.  But, the prefactor $\lv^{\alpha+\nu-1}$
diverges. Replacing $\rhohi(\lv z\tr)$ by $\rhohi(0)=1$, the integral can be
evaluated, and we find
\begin{equation}
  \label{plsmalls}
  \langle{(s+\rho)^{\alpha-1}} \rangle \sim \tr^{\nu}s^{\alpha+\nu-1}
  \frac{\Gamma(1-\alpha-\nu)}{\Gamma(1-\alpha)}, \quad
  \text{ for }\lv\tr \ll 1.
\end{equation}
Including the remaining factor of $s$ in the denominator, we see that
$\lt\Kda(s)\sim s^{-\alpha-\nu}$.  Due to the additional factor of $r$ in the
numerator of $\lt\Kra(s)$, this numerator does not diverge, but tends to
$\ensav{\rho^\alpha}$.  Thus, the reaction kernel
$\lt\Kra(s) \sim \ensav{\rho^\alpha} \tr^{-\nu} s^{- (\alpha + \nu)}$ and
$\lt\Kda(s) \sim \tr^{-\nu} s^{ - (\alpha + \nu)}$. These kernels provide the
Laplace-space definitions of the fractional derivative of order
$1-(\alpha + \nu)$ in~\eqref{nldr}.

\subsection{Random rate model for $\ensav{\tau}<\infty$ \label{supp:noheavytail}}

We have seen that for $\ensav{\tau}<\infty$, the system is perfectly well-mixed
in the scaling limit.  However, when $\ts>0$, the degree of mixing varies.
Details of the rate and waiting-time distributions appear in the survival
probability $p(t)$, and the solutions are rather complicated.

The following example offers a good illustration of how the system goes from
well-mixed to poorly-mixed as $\ts$ increases. We assume exponentially
distributed waiting times and $n$ discrete decay rates $\{r_{i}\}$, and find
that $p(t)$ decays as a sum of $n$ exponentials whose rates $\{b_{i}\}$ cannot
be easily computed in general. $p(t)$ decays at long times as the slowest of
these rates.  We assume exponentially distributed waiting times with density
\begin{equation}
\tpdf(t) = (1/\ts)\exp(-t/\ts),
 \label{expwtd}
\end{equation}
so that $\ensav{\tau} = \ts$.  Then $p(t)$ in Laplace space becomes
\begin{equation}
\lt p(s) = \frac{\ensav{\frac{1-\lttpdf(s + \rho)}{s+\rho}}}{\ensav{1- \lttpdf(s+\rho)}}
 = \ts \frac{ \ensav{\frac{1}{1+\ts(s+\rho)}}}{1 - \ensav{\frac{1}{1+\ts(s+\rho)}}}.
 \label{poftexplap}
\end{equation}
Note that in~\eqref{poftexplap}, \textit{no} approximations have been made.  The
density of discrete rates is given by
\begin{equation}
  \rho(\rv) = \frac{1}{n} \left[ \delta(\rv-\rv_{1}) + \delta(\rv-\rv_{2}) + \cdots + \delta(\rv-\rv_{n})  \right],
 \label{discrrate}
\end{equation}
where we have chosen equal weights for simplicity.
Substituting~\eqref{discrrate} into~\eqref{poftexplap}, we find
\begin{equation}
  p(\lv) = \frac{\frac{1}{1+(\lv+\rv_{1})\ts} +  \frac{1}{1+(\lv+\rv_{2})\ts} + \cdots}
    { \frac{\lv + \rv_{1}}{1+(\lv+\rv_{1})\ts} +  \frac{\lv + \rv_{2}}{1+(\lv+\rv_{2})\ts} + \cdots}
 \label{expwtddiscr}
\end{equation}
Putting all terms in the numerator over a a common denominator and likewise with
all terms in the denominator and then canceling the common denominator, we
arrive at a fraction with a polynomial of order $n-1$ in $\lv$ in the numerator
and a polynomial of order $n$ in $\lv$ in the denominator. Note that if we
rewrite the following expression with a common denominator
\begin{equation}
  \frac{a_{1}}{\lv + b_{1}} + \frac{a_{2}}{\lv + b_{2}} + \cdots  + \frac{a_{n}}{\lv + b_{n}}.
 \label{fexpression}
\end{equation}
we obtain again a fraction with a polynomial of order $n-1$ in the numerator and
$n$ in the denominator.  Because inverting this last expression gives a sum of
exponentials, inverting~\eqref{expwtddiscr} also gives a sum of
exponentials. (There can be no oscillating modes.)  The decay rates of $p(t)$
are found by equating coefficients in the two expressions~\eqref{expwtddiscr}
and~\eqref{fexpression}, with the result that $\{b_{i}\}$ are given by the roots
of the polynomial in $s$
\begin{equation}
  \sum_{j=1}^{n}  (\lv - \rv_{j}) \prod_{i\ne j} (1 + [\lv - \rv_{i}]\ts).
 \label{exppoly}
\end{equation}
Dividing~\eqref{exppoly} by $\ts^{n-1}$ and expanding about $1/\ts = 0$, we find
that for $\ts \gg 1/r_{j}$, $j=1,\ldots$, the effective rates are equal to the
discrete disordered rates $\{b_{i}\} = \{r_{i}\}$. This is the poorly-mixed and
highly-segregated limit. Each particle decays in its initial environment. At
long times, $p(t)$ decays exponentially at the rate equal to the smallest of
$\{r_{i}\}$.

In the opposite limit $\ts \ll 1/r_{j}$, $j=1,\ldots$, the
polynomial~\eqref{exppoly} has singular roots. The roots may be found by regular
perturbation after substituting $y=s\ts$ and multiplying by $\ts$, so
that~\eqref{exppoly} becomes, to leading order in $\ts$
\begin{equation*}
  \left(\ts \sum_{j=1}^{n} r_{j} - n y \right) (1-y)^{n-1}.
\end{equation*}
Thus, the roots are $s=\ensav{\rho}$, and the $(n-1)$-fold degenerate value
$s=1/\ts$.  This shows how the well-mixed limit of a single homogeneous rate
$\ensav{\rho}$ is approached with increasing $\ts$.  All modes except the
homogeneous mode decay rapidly. Only the homogeneous mode survives the scaling
limit $\ts\to\infty$, so the decay is purely exponential for all times.  For
$n=2$ the rates of the multi-exponential decay of $p(t)$ take the explicit form
\begin{equation}
  \frac{1}{2\ts} \left[1  +\ts(r_{1}+r_{2})  \pm \sqrt{\ts^2(r_{1}-r_{2})^2+1} \right],
 \label{exactexppsi}
\end{equation}
which shows that transport and decay are coupled in the intermediate regime.
The coefficients of the two terms corresponding to the rates~\eqref{exactexppsi}
are
\begin{equation*}
  \frac{1}{2}\left\{ 1 \mp \left[1 + \ts^{2}(r_{1}-r_{2})^{2} \right] \right\}.
\end{equation*}
We see that, as $\ts\to 0$ the coefficient corresponding to the rate that
diverges as $1/\ts$ tends to zero.

Finally, we note that the example above is the solution for the survival time of
the first passage time under reset process, with the density of the underlying
FPT given by~\eqref{discrrate}, and exponentially distributed reset time.

\subsection{Stochastic simulations \label{app:MC}}

We verify and illustrate the analytic results using stochastic simulations of
the microscopic model as presented in~\secref{sec:model}.
The analytic results are plotted using exact asymptotic expansions as well as
numerical inverse Laplace transform (ILT) of the solutions in Laplace space.
The simulation curves represent averages over $10^{6}-10^{8}$ trials.
Inverse Laplace transform is used in Figs.~\ref{fig:PandMSD},~\ref{fig:varyalphanu},~\ref{fig:propagator},
\ref{fig:msdsim}, and~\ref{fig:instantrate}. Dashed and dotted lines in these figures are obtained from
real-space asymptotic expansions. Stochastic simulations are used in~Figs~\ref{fig:varyalphanu}
and~\ref{fig:msdsim}. These figures clearly show excellent agreement between theoretical predictions
and simulations of the microscopic model.
The analytic results are mostly based on limiting forms of the PDFs for large or
small arguments.  For simulations, and ILT, we used the following concrete PDFs.
To represent the heavy-tailed waiting time density~\eqref{wtdheavy}, we choose
the Pareto distribution.
\begin{equation}
  \psi_{\tau}(t) =
  \begin{cases}
      \alpha \ts^{\alpha}/\Gamma(1-\alpha)t^{-\alpha-1}  &  \text{for } \   t > \ts \Gamma(1-\alpha)^{-1/\alpha}  \\
      0   &  \text{ otherwise }
  \end{cases}.
  \label{pareto}
\end{equation}
To represent the rate PDF~\eqref{rhopower} for~\figref{fig:PandMSD}, we used the upper-truncated density
\begin{equation}
  \psi_{\rho}(r) =
  \begin{cases}
      \nu r_{0}^{-\nu} r^{\nu-1}  &    r < r_{0} \\
      0   &  \text{ otherwise }
  \end{cases},
  \label{uppertrunc}
\end{equation}
where $r_{0} = \tr^{-1} \Gamma(\nu+1)^{1/\nu}$, and $\rhohi(z)\equiv 1$.  For
the other figures we instead used the gamma distribution, which is
exactly~\eqref{rhopower} with $\rhohi(z) = \exp(-z)$. In accordance with the
discussion leading to~\eqref{Dtpdfdef}, we also sampled from the PDF for $\Dt$
directly.  For the gamma distribution, the integral
$\psi_{\Dt}(t) = \ensav{\rho \exp(-\rho t)}$ takes the form
\begin{equation}
   \psi_{\Dt}(t) = \nu\tr^{-1} (1+t/\tr)^{-\nu-1}.
  \label{psigamma}
\end{equation}
It is well known that one can easily sample from a distribution if its
cumulative distribution function (CDF) can be inverted. The CDF corresponding
to~\eqref{psigamma} is $C(t) = 1 - (1+t/\tr)^{-\nu}$.  The functional inverse is
$t(C) = \tr[ (1-C)^{-1/\nu} -1 ]$. Samples of the PDF~\eqref{psigamma} are
obtained by substituting pseudo-random numbers uniformly distributed on $(0,1)$
for $C$ in $t(C)$.

The curves in~Fig.~\ref{fig:varyalphanu} that were obtained by numerical
inversion are based on~\eqref{psurvl}. The corresponding Monte Carlo (MC) curves
in the same figure were obtained by averaging $10^{6}-10^{7}$ trials of the
microscopic model.  The rates were sampled from a gamma distribution with mean
$1$, and a microscopic timescale $\ts=1/100$. Thus, systematic deviations of the
MC from the scaling limit~\eqref{psurvl} are not visible
in~\figref{fig:varyalphanu}.

\figref{fig:msdsim} verifies the expression for the MSD $m(t)$ given
by~\eqref{exactsecondmom} with waiting time PDF characterized by $\alpha=1/4$,
$\ts=0.1$, and two equally probable rates, $0$ and $\tr$.  represented by the
rate PDF~\eqref{tworates} with $p=1/2$. We show curves for two values of $\tr$,
$\tr=10^{3}$, and $\tr=10^{4}$.  The curves compare ILT with averages over
$1.5 \times 10^{7}$ simulations.  Due to statistical noise, it is difficult to
probe the long time behavior of $m(t)$ in the scaling limit with simulations.
Deviation from the scaling limit is clearly visible in~\figref{fig:msdsim}.  On
this double-log plot, the curves would be linear for $t < \tr$ in the scaling
limit. In particular, they would coincide with the scaling limit of $m(t)$ for
the non-reactive CTRW shown in~\figref{fig:PandMSD}.  To obtain agreement with
the simulations on the scale of this plot, it is sufficient to include the first
correction to the scaling limit of~\eqref{exactsecondmom} for the ILT. As
expected, the crossover to localized behavior clearly occurs for $t\approx \tr$.

\subsubsection{Survival probability.~~}
Here we present the method and algorithm used to compute an MC estimate of the
survival probability $p(t)$ from simulations.  The survival probability can be
written
\begin{align}
\label{survprobcdf}
 p(t) = 1 - \int_{0}^{t} p_{d}(t') \d{t'} = 1 - C_{d}(t),
\end{align}
where $p_{d}(t)$ is the probability density for the death time, and $C_{d}(t)$
is the cumulative distribution function (CDF) of the death time.  To make clear
the meaning of death time: the probability that a particle dies between time $t$
and $t+\d{t}$ is $p_{d}(t) \d{t}$, given that is alive and untrapped at time
$t=0$. We compute the empirical CDF corresponding to $C_{d}(t)$, which is an
unbiased estimator that converges to $C_{d}(t)$~\cite{Billingsley1995}.

The empirical CDF of the CDF $C(x)$ of a random variable $X$ is computed as
follows. 1) Draw $n$ independent samples of $X$, storing each one in an array
$A$ in sampling order. 2) Sort the array $A$ in increasing order. In particular,
after sorting, the first element $A_{1}$ is the least sample and the last
element $A_{n}$ is the greatest sample.  The empirical CDF is given by the
points $(A_{i},i/n)$. To be clear, the coordinate is $A_{i}$ and the ordinate is
$i/n$.  Referring to~\eqref{survprobcdf}, the empirical survival probability is
given by $(A_{i},1-i/n)$.

\subsubsection{Mean square displacement.~~}

We compute MC estimates as follows. The time increments of a particle's
trajectory and its death time are generated as in the previous section. However,
we also track the position of the walker at each step, using normally
distributed step displacements with unit variance. In order to perform an
ensemble average over trajectories, we must establish an array of fixed times
$t_{\text{rec}}$ at which to record the MSD. We also maintain an array
$m_{\text{rec}}$ of the same length. The element $m_{\text{rec},i}$ contains the
sum over particle trajectories of the squared displacement recorded at time
$t_{\text{rec},i}$.  We consider, as before, two cases. Either the walker dies
during a step, or does not. Consider the second case. The walker is at position
$x=\sum_{j=1}^{i-1} x_{j}$ during the time interval $(t,t+t_{i})$. We maintain
an index $i_{m}$ into the array $t_{\text{rec}}$ corresponding the most recent
time at which the squared displacement for this trajectory was recorded.  We
then check which of the recording times
$t_{\text{rec},i_{m}+1},t_{\text{rec},i_{m}+2},\ldots$ lie in the interval
$(t,t+t_{i})$. For each of the corresponding indices $i_{m}+1$, etc. we add
$x^{2}$ to the element $m_{\text{rec},i_{m}+1}$, etc.  We advance $i_{m}$ to the
last recorded time. We then proceed to the next step. Now consider the first
case, when the particle dies. The relevant time interval is now
$(t,t+\delta t)$, because the particle dies at $t+\delta t$. We only record
$x^{2}$ at recording times lying in this final interval. In both cases, we also
increment the number of particle trajectories $n_{i}$ contributing to the sum at
each recording time $t_{\text{rec},i}$. This number of course decreases with
increasing time because particles are dying. The estimate of the MSD normalized
by the survival probability is then $m_{\text{rec},i}/n_{i}$. Suppose the total
number of trials is $n$. The estimate of the MSD normalized by the total number
of particles, live or dead is of course $m_{\text{rec},i}/n$. The relation to
the analytic quantities is
\begin{align}
 \int_{-\infty}^{\infty} \d{x} \, x^{2} p(x,t)  \Leftrightarrow m_{\text{rec},i}/n,
\end{align}
and
\begin{align}
 \msd(t) = \frac{\int_{-\infty}^{\infty} \d{x} \, x^{2} p(x,t)}{\int_{-\infty}^{\infty} \d{x} \, p(x,t)}  \Leftrightarrow m_{\text{rec},i}/n_{i}.
\end{align}
As an example to understand the difference: Suppose only live particles are
detectable. Then $\msd(t)$ describes the observed width of the cloud. Note that
the corresponding estimate of $p(t)$ is $n_{i}/n$, so that we have, as expected
$(m_{\text{rec},i}/n)/(n_{i}/n) = m_{\text{rec},i}/n_{i}$. In practice, we
instead estimate $p(t)$ using the method described in the previous section,
which is far more efficient. We used the Mersenne Twister RNG.


\section*{Acknowledgments}
This work was supported by the European Research Council (ERC) through the
project MHetScale (Contract number 617511)


%

\end{document}